\documentclass[10pt,journal,compsoc]{IEEEtran}
\IEEEoverridecommandlockouts
\usepackage{cite}
\usepackage{amsmath,amssymb,amsfonts}
\usepackage{enumitem}
\usepackage{graphicx}
\usepackage{textcomp}
\usepackage{xcolor}
\usepackage{hyperref}       
\usepackage{url}            
\usepackage{booktabs}       
\usepackage{amsfonts}       
\usepackage{mathtools}
\usepackage{float}
\usepackage{dblfloatfix}    
\usepackage{multirow}
\usepackage{algpseudocode}
\usepackage[ruled,vlined,linesnumbered]{algorithm2e}

\usepackage{amsthm}
\usepackage{etoolbox}

\AfterEndEnvironment{theorem}{\noindent\ignorespaces}

\AfterEndEnvironment{corollary}{\noindent\ignorespaces}

\AfterEndEnvironment{lemma}{\noindent\ignorespaces}
\newtheorem{definition}{\textbf{Definition}}
\AfterEndEnvironment{definition}{\noindent\ignorespaces}

\AfterEndEnvironment{example}{\noindent\ignorespaces}
\usepackage{amssymb}
\usepackage{amsmath}

\DeclareMathOperator*{\argmin}{arg\,min} 

\usepackage{xfrac}
\usepackage{array}
\newcolumntype{x}[1]{>{\centering\arraybackslash\hspace{0pt}}p{#1}}

\def\BibTeX{{\rm B\kern-.05em{\sc i\kern-.025em b}\kern-.08em
    T\kern-.1667em\lower.7ex\hbox{E}\kern-.125emX}}
\begin{document}

\title{A Kernel Method to Nonlinear Location Estimation with RSS-based Fingerprint
}

\author{Pai Chet Ng, ~\IEEEmembership{Member,~IEEE,} 
	Petros Spachos,~\IEEEmembership{Senior Member,~IEEE}
	James She,~\IEEEmembership{Member,~IEEE} and
	Konstantinos N. Plataniotis,~\IEEEmembership{Fellow,~IEEE}
	\thanks{Pai Chet Ng is with the Department of Electrical and Computer Engineering, University of Toronto, Canada. E-mail: pc.ng@utoronto.ca}
		\thanks{Petros Spachos is with the School of Engineering, University of Guelph, Canada. E-mail: petros@uoguelph.ca}
		\thanks{James She is with the Division of Information and Computing Technology, College of Science and Engineering, Hamad Bin Khalifa University, Qatar. E-mail: pshe@hbku.edu.qa}

	\thanks{Konstantinos N. Plataniotis is with the Department of Electrical and Computer Engineering, University of Toronto, Canada. E-mail: kostas@ece.utoronto.ca}
}

\IEEEtitleabstractindextext{
\begin{abstract}
This paper presents a nonlinear location estimation to infer the position of a user holding a smartphone. We consider a large location with $M$ number of grid points, each grid point is labeled with a unique fingerprint consisting of the received signal strength (RSS) values measured from $N$ number of Bluetooth Low Energy (BLE) beacons. Given the fingerprint observed by the smartphone, the user's current location can be estimated by finding the top-k similar fingerprints from the list of fingerprints registered in the database.
Besides the environmental factors, the dynamicity in holding the smartphone is another source to the variation in fingerprint measurements, yet there are not many studies addressing the fingerprint variability due to dynamic smartphone positions held by human hands during online detection. 
To this end, we propose a nonlinear location estimation using the kernel method. Specifically, our proposed method comprises of two steps: 1) a beacon selection strategy to select a subset of beacons that is insensitive to the subtle change of holding positions, and 2) a kernel method to compute the similarity between this subset of observed signals and all the fingerprints registered in the database. The experimental results based on large-scale data collected in a complex building indicate a substantial performance gain of our proposed approach in comparison to state-of-the-art methods. The dataset consisting of the signal information collected from the beacons is available online.
\end{abstract}

} 
\maketitle
\IEEEpeerreviewmaketitle

 \section{Introduction}
\label{sec:intro} 
\IEEEPARstart{L}{ocation} estimation is essential to many smartphone applications targeting context-aware services in indoor environments~\cite{6985718, 9094357}. 
Most of these applications exploit wireless signals to passively estimate the user's location in the background while having the context-aware services run in the foreground.
There are two common approaches to wireless-based location estimation: direct approach and indirect approach. The direct approach is a single-step estimation that estimates the location by directly processing the wireless signals; whereas the indirect approach involves an intermediate step in computing the distance, time of arrival (ToA), or angle of arrival (AoA) prior to location estimation.
The computed distance, ToA, and AoA are then used as input parameters by trilateration or triangulation algorithms to estimate the target location~\cite{7945521, 8371230, 9036937, 5427001, 9257105}.
Recently, location estimation based on the direct approach has attracted a lot of interest because it can estimate the location directly without going through the intermediate step.
The direct approach is achieved by exploiting wireless signals from densely deployed transmitting and receiving sources. For example,~\cite{9098064, 7500062} use a large-scale antenna array and~\cite{6399619} uses the received signal from multiple base stations to achieve direct localization.

Location estimation based on fingerprinting is a direct approach that can estimate the location directly given the observed signals.
The location is estimated by computing the similarity of the signals observed during the online stage to the list of fingerprints registered during the offline stage.
Specifically, the fingerprint-based location estimation can be divided into two stages: 1) offline stage and 2) online stage~\cite{6059451}.
During the offline stage, a series of on-site surveys are performed to register the fingerprint at a particular grid point in a given location.
A fingerprint database is used to store all the registered fingerprints, and each grid point can be represented by a unique fingerprint vector~\cite{7103024}.
In this paper, we use the ambient Bluetooth signal for constructing the fingerprint vector $\mathbf{f} \in \mathbb{R}^N$.
Each entry of $\mathbf{f}$ denotes the time average received signal strength (RSS) measured from a Bluetooth Low Energy (BLE) beacon.
The reason to use BLE beacons is that BLE is ubiquitous and the RSS value can be measured easily with any modern smartphone~\cite{8242361, 9001059}.
Given a system with $N$ number of BLE beacons and $M$ number of grid points, the database should consist of $M$ number of unique fingerprint vectors $\mathbf{f}_i \in \mathbb{R}^N$ for all $l_i \in L$, where $L$ is a set of location grid points and $|L| = M$.

\begin{figure}
	\centering
	\includegraphics[width=1\columnwidth]{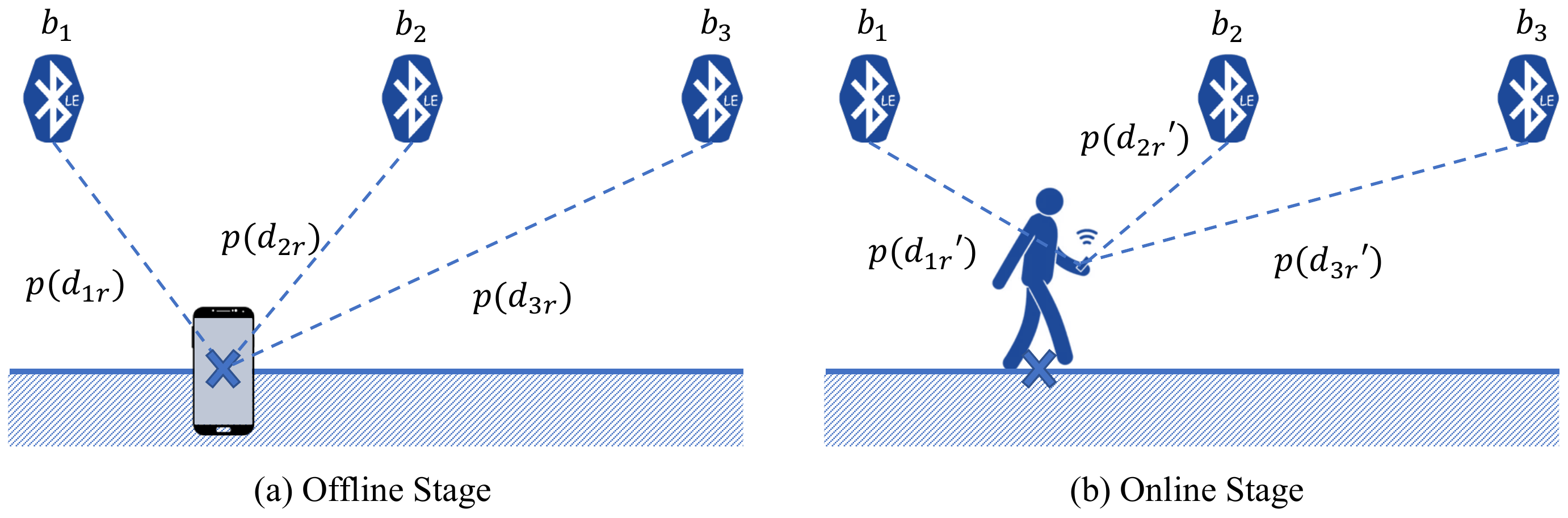}
	\caption{During the on-site fingerprint survey, the receiver is placed stationary on a grid point (e.g., on the floor or a non-moving robot's arm) for a certain duration. However, during the online stage, different users might hold their smartphones differently even though they are at the same grid point.}
	\label{fig:estProblem}
\end{figure}
 
During the online stage, the location of a user can be estimated by computing the similarity between the observation vector $\mathbf{o} \in \mathbb{R}^N$ and the list of fingerprints registered in the database.
Such a direct similarity computation cannot guarantee a good performance owing to the discrepancy between the registered fingerprint and the current observation.
Besides environmental factors, another major factor leading to such a discrepancy is how the user holds her/his smartphone during online detection.
This is a real problem for smartphone-based location estimation aiming for delivering context-aware services because the user might hold the smartphone with different poses and gestures while interacting with the context-aware service provided by the smartphone. 

Consider the scenario during the on-site survey, as described in Fig.~\ref{fig:estProblem}(a), the receiver will be placed stationary during fingerprint registration so that it can obtain a more robust representation~\cite{4161920}. 
For example, we can put the receiver on the robot arm and the robot arm will remain stationary during the registration process. 
We can also have the receiver held by a tripod or placed on the floor.
However, it is impossible for the user to hold her/his smartphone the same way the receiver was placed during the on-site survey, not to mention that the user does not even have the knowledge about the process of the on-site survey.
From Fig.~\ref{fig:estProblem}(b), we can see that the smartphone is always held at different heights from the initial position where the fingerprint is registered.
So, even though the user stands on the exact same grid point where the fingerprint is registered, the observation vector always looks dissimilar to the fingerprint stored in the database. 
While this subtle difference in distance from the initial position seems negligible, it was observed that a small change in distances may result in a very different RSS value even though the user remains still at the same grid point.
While many works have studied the variation induced by the environmental factors (e.g., multipath~\cite{7174982, 7488215} and shadowing effects~\cite{5071219 ,8588347}), there are no work studies such a discrepancy.

Let $d_{ir}$ be the distance between the $i$th beacon to the receiver $r$, the RSS value measured by the receiver at $d_{ir}$ would be $p(d_{ir})$, where $p(\cdot)$ is a function that returns the RSS value. 
One model that has been widely used to estimate $p(\cdot)$ is the path-loss model.
From Fig.~\ref{fig:estProblem}~(a) and (b), it is clear that $d_{ir} \neq d_{ir}', \forall i \in B$.
Such a subtle change in distances might cause a large variation in the RSS measured by the smartphone. 
However, our empirical analysis unveils that not all the beacons will give a very different RSS value due to such a small change in distances.
Most surprisingly, we discovered that the strongest RSS value measured from the nearest beacon is the one that suffers the most variation.
On the contrary, those beacons located at farther locations can produce more robust RSS measurements despite the dynamic holding position of smartphone.
In light of such observations, we propose a beacon selection strategy to only include those RSS values that are more robust to the subtle change in distances.
More specifically, our proposed method manipulates weak signals from farther beacons during the online detection stage to estimate the location rather than the strongest signal which has been exploited in many works~\cite{abed2019rss, 6846747, ezhumalai2021efficient}.

Given the observation vector $\mathbf{o} \in \mathbb{R}^{N'}$ refined by our beacon selection strategy, we can then estimate the location by performing the similarity computation. 
One common approach is to use the $k$ nearest neighbors (kNN) to search for the top $k$ similar fingerprints, and then apply the weighted average to compute the location~\cite{8993814}. 
However, the kNN method always returns incorrect location estimation due to the linearly non-separable fingerprint data in $\mathbb{R}^{N'}$.
To this end, we employ the kernel method to compute the similarity between the fingerprint vector in the database and the observed fingerprint in a high-dimensional $H$-space without literally transforming the fingerprint vectors into the $H$-space.
Conventionally, one would map the linearly non-separable fingerprint data $\mathbf{f} \in \mathbb{R}^N$ to a high-dimensional $H$-space such that the data becomes linearly separable, before proceeding with the similarity computation.
Rather than going through the above two steps, we define a kernel function to compute the similarity between these high-dimensional data $\mathbf{f} \in \mathbb{R}^H$ without literally visiting the $H$-space. 
The main contributions of this paper are:
\begin{itemize}
	\item Beacon selection strategy: Based on our experimental observations, it unveils that the strongest signal from the nearest beacon is the one that suffers the most RSS variation even though the user remains still at the same location. Those farther beacons, on the other hand, even though contribute a  weaker signal, their RSS values are relatively robust regardless of the small change in distances induced by hand's movement. In light of such empirical argument, we propose a novel selection strategy to only include the RSS values from the beacons that are insensitive to the small change in distances. 
	\item Extensive experiments: our experiments show that our beacon selection strategy can further improve the localization performance, achieving sub-meter accuracy compared to the state-of-the-art method. In particular, the localization error is reduced from an average of 3.5~m to less than 1~m, when applying our selection strategy to the conventional kNN method.
	\item Large-scale RSS dataset: We performed a series of real experiments in a complex building to verify the performance of our proposed approach. The outcome of the experiments produces the first large-scale RSS dataset that consists of 100k training samples and 60k testing samples. The dataset is made publicly available to encourage further research.
\end{itemize}
Note that this paper mainly focuses on the online location estimation assuming the fingerprint data is readily available. 
There are many works that discuss the problem related to the offline on-site survey. Interested readers can refer to the following fingerprint surveying methods: human-labor~\cite{6081874}, crowdsourcing~\cite{6805641}, semi-auto fingerprint generating~\cite{yang2012locating} and simultaneous localization and mapping~\cite{8057286, 8920098}.

The rest of the paper is organized as follows.
Section~\ref{sec:relatedWork} succinctly review the state-of-the-art methods in location fingerprint.
Section~\ref{sec:lf} provides the overview of location fingerprint describing the pre-processing in the offline stage and post-processing in the online stage.
Section~\ref{sec:db} includes the proposed beacon selection strategy.
Section~\ref{sec:ea} supports the proposed method with thorough empirical analysis.
Section~\ref{sec:ex} presents our experimental testbed used for data collection and describes our collected data.
Section~\ref{sec:ex2} evaluates the localization performance in comparison to selected baseline methods.
Section~\ref{sec:conclusions} concludes the paper with future works.

\begin{table}
	\caption{The choice of wireless signals and computation techniques}
	\label{table:relWork}
	\centering
	\begin{tabular}{l|p{1.9cm}p{2.5cm}}
		\toprule
		\cmidrule{1-3}
		Work &Choice of wireless signals &Computation techniques \\
		\hline
		Yoon et al. \cite{7181720}            &FM              & Euclidean distance\\
		Chen et al. \cite{6516864}            &FM              & kNN\\
		Wang et al. \cite{wang2013dude}       &RFID            & Dynamic time warping\\
		Yang et al. \cite{yang2014tagoram}    &RFID            & Differential augmented hologram\\ 
		Tarzia et al. \cite{tarzia2011indoor} &Acoustic        & kNN\\
		Bahl et al. \cite{832252}             &WiFi            & kNN\\
		Caso et al. \cite{8680659}            &WiFi            & WkNN\\
		Luo et al. \cite{8693854}             &WiFi            & LDA\\
		Hoang et al. \cite{8830368}           &WiFi            & RNN\\
		Ouyang et al. \cite{6018966}          &WiFi            & Expectation-maximization\\
		Sun et al. \cite{8464281}             &WiFi            & Gaussion Process\\
		Faragher et al. \cite{7103024}        &BLE             & kNN\\
		Ng et al.\cite{9000599}               &BLE             & Compressive sensing\\
		Song et al. \cite{7986446}            &WSN             & SVM\\
		\bottomrule
	\end{tabular}
\end{table}
\section{Related Work}
\label{sec:relatedWork}
Location estimation with wireless signals has been a hot topic from the 3G era until the 6G era. Recent works have been proposed to integrate 6G communication and sensing together for outdoor localization~\cite{bourdoux20206g, xiao2020overview}.
Compared to outdoor localization, indoor localization is far more challenging owing to unpredictable environmental factors. 
Currently, RSS-based fingerprint methods have shown a substantial performance improvement for location estimation in indoor environments.
Two major components we need to consider when adopting RSS-based fingerprints are 1) the choice of wireless signals for fingerprinting and 2) the computation technique to estimate the location.
Table~\ref{table:relWork} summarizes the choice of \textit{wireless signals} and \textit{computation techniques} adopted by related work.
Accordingly, this section first reviews the related wireless signals for fingerprinting and then discusses the current development of machine learning methods for location estimation.

\subsection{Ambient Wireless Signals for Fingerprinting}
Many works have exploited wireless signals, for example, FM radio~\cite{7181720, 6516864}, RFID~\cite{wang2013dude, yang2014tagoram}, acoustic~\cite{tarzia2011indoor},  WiFi~\cite{832252, 8680659}, and BLE~\cite{7103024, 9000599} for fingerprinting. 
Among them, WiFi access points are the popular choice since they are readily available in many public locations; whereas the ubiquity of Bluetooth technologies introducing new RSS-based fingerprinting with BLE beacons.
Other wireless technologies, on the other hand, require specific infrastructures and dedicated devices, which are not always available to end-users.
Since our work focuses on location estimation with smartphones, this section mainly reviews the wireless signals (i.e., WiFi and Bluetooth) that are familiar to end-users.

\subsubsection{WiFi Access Points}
Leveraging the pervasive deployment of existing WLAN infrastructures~\cite{7174948}, many works exploit RSS values from WiFi access points for fingerprinting~\cite{4161920}, including RADAR~\cite{832252}, Horus~\cite{youssef2005horus}, and EZPerfect~\cite{li2014experiencing}.
While promising results have been achieved, none of the above works examine the problem related to the varying heights of a smartphone subject to how the user holds the smartphone.
For example, RADAR only considers the different positions with respect to the location grid points; whereas EZPerfect only considers the RSS problem due to different device types.

WiFi access points are always installed according to a deployment plan to ensure wide coverage in popular and crowded areas.
Some works leverage this deployment information to improve localization accuracy.
For example, MapCraft~\cite{6846747} uses the signal from the strongest access point as a landmark and constrains the location estimation to a subset of access points.
HALLWAY~\cite{jiang2013hallway}, on the other hand, uses a subset of access points, in which their signal strengths are in ascending order, to first classify a coarse-grain area prior to estimating a fine-grain location.
The problem with this deployment planning is that it aims to maximize the coverage at those crowded locations, while ignoring some areas with fewer visits, such as staircases, underground garages, etc.

\subsubsection{BLE Beacon}
While BLE is a short-range and low-power technology mainly aiming for Internet of Things (IoT) development~\cite{8419192}, the accessibility of BLE signals, even in those less-visited areas, has made BLE an alternative choice for RSS fingerprinting.
Many fingerprint-based localization techniques leverage the broadcasting feature of BLE beacons to estimate the location of a user holding a smartphone~\cite{7103024, 7959171, 7827145}. 
The work in~\cite{7103024} exploits the RSS values measured from all the deployed beacons to construct a location fingerprint.
The work in~\cite{7959171} uses a deep learning approach to learn a robust fingerprint representation for 3D localization.
On the other hand,~\cite{7827145} computes the proximity information based on the Gaussian process regression to provide loosely location information.
Overall, none of these works consider the fingerprint variability due to the dynamic holding gesture by different users.

Since Bluetooth is a ready technology in most smartphones, it makes BLE a suitable wireless choice when considering an indoor localization involving a smartphone.
However, most BLE devices are powered by a coin-cell battery and might stop functioning from time to time. 
Even though~\cite{8705339} leverages the compressive sensing approach to reconstruct the fingerprint prior to location estimation, the reconstructed fingerprint still could not address the problem related to the holding variations of a smartphone. 
The massive deployment of BLE beacons in many indoor locations, on the other hand, resulted in a high-dimensional fingerprint vector and thus a large amount of data could be collected during the on-site survey.
To reduce the amount of data,~\cite{9075151} uses a traditional trilateration approach to estimate the initial location before correcting the location estimation using a light-weight fingerprint map. 
However, such a lightweight fingerprint map is only used to correct the initial estimation error without considering which beacons should be used to refine the fingerprint.

In our work, we propose a beacon selection strategy to identify which beacons are less sensitive to the holding variations so that a better location estimation can be achieved while reducing the size of the fingerprint.

\subsection{Location Estimation with Machine Learning}
Many machine learning methods have been used to improve the location estimation with RSS fingerprints. Most of them can be grouped into two types: supervised learning and unsupervised learning. 
Supervised learning is used to train a model to estimate the location given a set of labeled data; whereas unsupervised learning is normally used to learn some clusters or perform noise filtering given a set of unlabeled data.

\subsubsection{\textbf{Supervised Learning}}
Many localization problems leveraged supervised learning, requiring a labeled dataset associating each fingerprint vector to a labeled grid point,  to train a classifier to classify the location.
The $k$ nearest neighbor (kNN) is the commonly used classification method for many fingerprint-based localizations, including the fingerprint-based on WiFi~\cite{832252} and beacon~\cite{7103024}.
Besides kNN, support vector machine (SVM)~\cite{7986446},  linear discriminant analysis (LDA)~\cite{8693854}, and neural networks (NN)~\cite{8792196} have also been widely exploited to improve localization performance.
All these classifiers are trained via a deterministic approach by explicitly mapping the input fingerprint vector to the associated output label.
Such approaches suffer severe performance degradation when RSS values in the fingerprint vector change due to unpredictable environmental factors.

Besides those deterministic approaches, some probabilistic approaches, including Bayesian network~\cite{nandakumar2012centaur}, expectation-maximization~\cite{6018966},  Gaussian process~\cite{8464281}, have been used to estimate the grid point that returns the maximum likelihood.
Based on the statistical distribution of the RSS value at each grid point, the probabilistic approach can infer the location based on the confidence interval.
While these probabilistic approaches are more flexible to the changing environment, this kind of method always relies on certain assumptions to define the RSS distribution.
However, the RSS value may not always follow the same distribution pattern in a practical environment.

\subsubsection{\textbf{Unsupervised Learning}}
Unsupervised learning is always used to learn a better fingerprint matrix (commonly known as radio map) through clustering or noise filtering. 
For example, WiGEM~\cite{goswami2011wigem} learns the signal propagation parameters through the Gaussian mixture model (GMM) and then clusters each location based on the predicted signal strength.
On the other hand,~\cite{7414026} exploits the kMean approach to cluster RSS values before learning to classify the location through supervised learning.
UMLI~\cite{6554890} aims to reduce the computation and radio map construction by clustering the grid points with similar RSS patterns.
Since the constructed radio map highly relies on the clustered grid points, a small change in the deployment topologies might cause a large change in the RSS measurements at different clusters.
Hence, heavy retraining is often required to deal with the change in deployment topology.
Besides learning from the raw data, current works by~\cite{8827486} and~\cite{8264804} learn the soft information from the data rather than using the range or distance information for location estimation.
Since the extracted soft information contains all possible positional information regarding a node, future work can consider integrating the soft information for clustering.

Noise filtering through unsupervised learning can learn a robust fingerprint representation prior to location estimation. 
The work in~\cite{8766808} proposes a denoising-contractive autoencoder to learn the RSS-based fingerprint that captures the variation induced by the human body while filtering out the noise contributed by the environmental factor.
On the other hand,~\cite{7959171} mainly focuses on getting rid of any noise present in the fingerprint vector through a deep autoencoder so that it can achieve a finer location estimation with cleaner representation. 
The work in~\cite{8052139} defines a walking step model to automatically calibrate the fingerprint vector and at the same time filter out the possible noise through a deep neural network.
Even though noise filtering is essential in learning a better fingerprint representation, it is still unable to learn the fingerprint representation that is invariant to the holding position of a smartphone.
Note that some noise, for example, those induced by multipath effects, is useful information to enhance the localization.
Rather than applying any unsupervised learning methods to perform noise filtering, our work directly selects those beacons with the least RSS variation in connection to the dynamic holding position, while retaining those noise information for localization training.

\begin{table}
	\caption{Summary of Mathematical Notations}
	\label{table:notation}
	\centering
	\begin{tabular}{lp{6cm}}
		\toprule
		\cmidrule{1-2}
		Notation     &  description  					\\
		\midrule
		$\mathbf{f} \in \mathbb{R}^N$ & fingerprint vector in the input space \\
		$\mathbf{f} \in \mathbb{R}^H$ 	 &  fingerprint vector in a transformed space      \\
		$\mathbf{o} \in \mathbb{R}^N$ 	 &  observation vector      \\
		$\mathbf{l} \in \mathbb{R}^D$ 	 &  location vector representing the coordinate of a grid point      \\
		$N$ 	 & total number of BLE beacons      \\
		$M$ 	 & total number of grid points      \\
		$B$  &  a set of beacons \\
		$L$  &  a set of location grid points \\
		$\mathcal{R}_{ji}$  & a set of instantaneous RSS values from beacon $b_j$, acquired by the receiver located at $l_i$ over a scanning duration $T_d$ \\
		$\mathbb{L}$ & a $k \times D$ location matrix \\
		$T_a$    & advertising interval \\
		$T_s$    & scanning interval \\
		$T_d$    & scanning duration \\
		$d_{ji}$ & distance between the receiver at the grid point $l_i$ and the beacon $b_j$ \\		
		$s$      & number of selected beacons \\
		\bottomrule
	\end{tabular}
\end{table}

\section{Overview of Location Fingerprint}
\label{sec:lf}
This section presents the location estimation system based on location fingerprint and formulates the problem. Table~\ref{table:notation} has a summary of the mathematical notations that are used.

\subsection{Location Fingerprint with BLE Beacons}
The location fingerprint is a unique representation that can be used to label a grid point in a given location.
Let $L = \{l_1, l_2, \dots, l_M\}$ be a set of grid points for a given location, then the fingerprint database should have $M$ number of fingerprint vectors.
In this paper, we focus on the location fingerprint that uses the RSS values measured from the BLE beacon network.

\begin{definition} \textbf{(Location Fingerprint)}
	A location fingerprint is a unique representation of a grid point, where the similarity between any two fingerprints of two different grid points is greater than or equal to zero, but less than one.  
	\begin{equation}
	0 \leq \beta(\mathbf{f}_i, \mathbf{f}_j) < 1, \quad \forall j \neq i \in L
	\end{equation}
	where $\beta(\cdot)$ is the function that computes the similarity, and $\mathbf{f}_i$ and $\mathbf{f}_j \in \mathbb{R}^N$ are the fingerprint vectors at grid point $l_i$ and $l_j$, respectively. 
	The function $\beta(\cdot)$ outputs one when the fingerprint is compared to itself, i.e., $\beta(\mathbf{f}_i, \mathbf{f}_i) = 1$.
	\label{def:LF}
\end{definition}

The fingerprint-based location estimation system is categorized into two stages: 1) the offline stage involving an on-site fingerprint survey, and 2) the online stage focusing on real-time location estimation.
The following subsections describe the feature representation with RF signals collected during the on-site fingerprint survey and then formulate the location estimation problem.

\subsubsection{On-site Fingerprint Survey During Offline Stage}
Suppose that there are $N$ number of beacons, i.e.,  $B = \{b_j | 0 < j \leq N, j \in  \mathbb{Z}^+ \}$, then the fingerprint vector at $l_i$ can be described as follows:
\begin{equation}
	\mathbf{f}_i = 
	\begin{pmatrix}
	f_1  & f_2  & \cdots & f_j & \cdots & f_N 
	\end{pmatrix}^T
	\label{eq:fg}
\end{equation} 
where $f_j$ denotes the RSS value measured from beacon $b_j$. 
Recall from Fig.~\ref{fig:estProblem}, we define a function $p(d_{ji})$ that returns the RSS value subject to the distance between the receiver at the grid point $l_i$ and the beacon $b_j$. 
Hence,  each entry in $\mathbf{f}_i$ can be obtained as follows:
\begin{equation}
	f_j^{(l_i)} = \mathbb{E}( p(d_{ji}) )
	\label{eq:fj}
\end{equation}
For simplicity, the superscript $(l_i)$ is omitted when describing the whole fingerprint vector, as in Eq.~(\ref{eq:fg}).

\subsubsection{Location Estimation During Online Stage}
\label{ss:locationEstimation}
The main goal of location estimation is to define a mapping function that maps the online observation $\mathbf{o} \in \mathbb{R}^N$ to a $D$-dimensional spatial position describing a grid point, i.e., $\phi(\cdot) : \mathbb{R}^N \rightarrow \mathbb{R}^D$, where $D$ can be either 2 or 3. 
Considering that a user carrying a smartphone is more concerned about their $x$ and $y$ coordinates rather than the height of their smartphone, we only consider a 2-dimensional grid point, i.e., $l_i = (x, y)$ in this paper.
In the context of 3-dimension, we are more interested to know on which floor the user is rather than the height of the user at a particular location. 
In general, most of the current 3D localizations are referring to multi-floor localization, with the floor being the third dimension instead of the explicit height~\cite{8693854, 7913667}.

Even though the height of the smartphone is not our major consideration, it has a significant effect on the RSS measurements.
The way each user holds the smartphone varies according to their height and holding gesture, resulting in a few centimeters difference in terms of the distance between the smartphone and all the beacons.
However, a few centimeters change in distances may lead to location estimation error when the smartphone, by measuring the change in RSS, thought that the user has moved from one location to another, while, in fact, the user just slightly adjust their holding gesture.
While the nearest beacon that can provide the strongest RSS measurement is more reliable, it is very sensitive to a small change in distances,  which  may result in a big variation in RSS measurements.
We further provide an empirical analysis to support the above argument in Section~\ref{sec:ea}.

\subsection{Problem Formulation}
\label{sec:pf}
Given the observation vector $\mathbf{o} \in \mathbb{R}^N$ containing the instantaneous RSS values measured by the smartphone at any arbitrary time instant, the objective is to find the top-$k$ matches fingerprints and then retrieve the corresponding grid points' information to estimate the current location. Upon retrieving the grid points' information, the location can be estimated as follows:
\begin{equation}
\tilde{\mathbf{l}} = \mathbf{w}^T \mathbb{L}
\label{eq:estLoc}
\end{equation}
where $\mathbf{w} \in \mathbb{R}^{k}$ is the weight vector and $\mathbb{L} \in \mathbb{R}^{k \times D}$ is a location matrix defined based on a subset of location grid points $L_k \subset L$ whose fingerprints are very close to  $\mathbf{o}$. The size of the estimated location $\tilde{\mathbf{l}}$ is determined by the parameter $D$. 
While we only consider the case with $D = 2$, as explained in Section~\ref{ss:locationEstimation}, the equation defined above can be extended to any size of $D$. 
Note that if $w_i = 1/k$ for all $w_i \in \mathbf{w}$, then the estimated location is obtained by averaging the top-$k$ location grid points.

Clearly, the location can be estimated once a subset of location grid points $L_k$ that defines the location matrix $L$ is found. Hence, the problem is to find the index of top-$k$ matches fingerprints. If $k = 1$, the problem is reduced to find a location grid point from $L$ that produces the highest similarity score, i.e.,
\begin{equation}
[v, \mathbf{l}_i] = \max_{L} \beta (\mathbf{f}_i, \mathbf{o})
\label{eq:setL}
\end{equation}
where $\mathbf{l}_i$ is the corresponding grid point with the highest similarity score $v$, and $\beta (\cdot)$ is the function that computes the similarity. 
Similarly, the top-$k$ location grid points can be obtained based on the top-$k$ similarity scores returned by $\beta (\cdot)$.
Hence, the fundamental problem with location estimation is the similarity computation, which is described in the next subsection.

\subsubsection{Similarity Computation}
\label{ssec:sc}
The function $\beta (\mathbf{f}_i, \mathbf{o})$ computes the similarity between the observation vector $\mathbf{o}$ and all the fingerprint vectors $\mathbf{f}_i$ in the database. 
In our context of the location estimation, the function $\beta (\mathbf{f}_i, \mathbf{o})$ should be a decreasing function describing the relationship between $\mathbf{o}$ and all $\mathbf{f}_i$ in the database.
Hence, $\beta (\mathbf{f}_i, \mathbf{o})$ can be described with the following properties:
\begin{itemize}
	\item The output of $\beta (\mathbf{f}_i, \mathbf{o})$ is confined to $[0, 1]$.
	\item The function $\beta (\mathbf{f}_i, \mathbf{o})$ decreases monotonically subject to the distance between the smartphone and the beacon, i.e., 
	\begin{equation}
	d_{pi} \geq d_{qi} \implies \beta (\mathbf{f}_i, \mathbf{o}_p) \leq \beta (\mathbf{f}_i, \mathbf{o}_q)
	\end{equation}
	and
	\begin{equation}
	d_{pi} \geq d_{pj} \implies \beta (\mathbf{f}_i, \mathbf{o}_p) \leq \beta (\mathbf{f}_j, \mathbf{o}_p)
	\end{equation}
	where $\mathbf{o}_p$ and $\mathbf{o}_q$ denote the observation vector when the smartphone is located at the grid point $l_p$ and $l_q$, respectively.
\end{itemize}
We can use any distance metrics to define the similarity computation function as long as they satisfy the properties listed above. 

\subsubsection{Distance Metrics}
\label{ss:distMetrics}
One common choice of distance metrics for similarity computation is the cosine similarity, which can be described as follows:
\begin{equation}
\beta (\mathbf{f}_i, \mathbf{o}) = \frac{\langle \mathbf{f}_i, \mathbf{o} \rangle }{||\mathbf{f}_i|| \, ||\mathbf{o}||}
\label{eq:cosine}
\end{equation}
However, cosine similarity is an ineffective function when dealing with the RSS variations measured from the same set of beacons.
For example, when any two fingerprints are constructed based on the same set of beacons, said $b_1, b_3, b_7$, and the observation vector also consists of the RSS values from these three beacons, the cosine similarity will say that the observation vector is very close to these two fingerprints, regardless of the magnitude (i.e., the RSS value) measured from these three beacons.

On the other hand, Euclidean distance is a more effective function since it considers the distance between the two vectors instead of the angular measurement. 
However, Euclidean distance is a positively increasing function with respect to the distance between the smartphone and the beacon. That is, the metric will return a higher value when the distance between the smartphone and the beacon increases. 
Also, the output may range from zero to infinity. 
To satisfy the properties defined in Section~\ref{ssec:sc}, the Euclidean distance need to be scaled to the unit norm, and then the normalized distance has to be subtracted with 1 to confine the output range to zero and one.
The similarity computation based on the inverse-normalized Euclidean distance can be described as follows:
\begin{equation}
\begin{aligned}
\beta (\mathbf{f}_i, \mathbf{o}) &= 1 - \left\|  \frac{\mathbf{f}_i}{|\mathbf{f}_i|} - \frac{\mathbf{o}}{|\mathbf{o}|} \right\|_2 
= 1 - \sqrt{\sum_{i = 1}^{n} (\frac{f_j}{|\mathbf{f}_i|} - \frac{o_j}{|\mathbf{o}|})^2}
\end{aligned}
\end{equation}
where $||\cdot||_2$ computes the $l_2$ norm of a vector.
While $l_2$ norm is useful because of its convexity feature, it may capture unwanted noise during similarity computation due to the inherent nature of the $l_2$ norm ball. 
A better alternative to the Euclidean distance is the Cityblock distance that relies on the $l_1$ norm. 
Again, we need to perform normalization and subtraction by one to the original Cityblock distance.
Mathematically, the similarity computation based on the inverse-normalized Cityblock distance can be described as follows:
\begin{equation}
\begin{aligned} 
\beta (\mathbf{f}_i, \mathbf{o}) &= 1- \left\|  \frac{\mathbf{f}_i}{|\mathbf{f}_i|} - \frac{\mathbf{o}}{|\mathbf{o}|} \right\|_1 
= 1 - \sum_{j = 1}^{n} \left| \frac{f_j}{|\mathbf{f}_i|} - \frac{o_j}{|\mathbf{o}|} \right|
\end{aligned}
\end{equation}
where $f_j$ and $o_j$ are the $j$-th element belonging to vector $\mathbf{f}_i$ and $\mathbf{o}$, respectively. $n \leq N$ is the number of beacons observed by the smartphone.
We also show in our experiment (Section~\ref{sec:ex}) that the similarity score obtained with Cityblock distance achieves the best performance when the size of the observation vector $n$ is sufficiently large. In other words, the dimensional of the feature space is large enough to linearly separate each fingerprint in the database.
However, we argue that in Section~\ref{ss:locationEstimation}, not all the beacons provide robust RSS values in connection to the dynamic holding position of a smartphone. 
This justifies the reason to exclude some beacons and selectively use more robust RSS measurements for similarity computation.

\section{Proposed Solution}
\label{sec:db}
Node selection has been a widely investigated topic: \cite{6031934} casts the node selection problem into a combinatorial optimization problem, \cite{1603409} and \cite{1603410} rely on either the global or local knowledge to select the best node for sensor localization, and \cite{5555901} selects the node by distinguishing the node that receives the Line of Sight (LOS) signal from the node receiving Non Line of Sight (NLOS) signals. 
So far, none of these works consider the node selection problem due to the sensitivity of the RSS values in connection with the small change in distances (less than a few centimeters). While such sensitivity might beneficial to those works aiming to achieve sub-centimeter accuracy, it is one of the factors causing location estimation error. 
Supported by our empirical findings in the next section, this section proposes our novel beacon selection strategy. While our beacon selection strategy might affect the performance of the similarity computation based on the distance metrics described in the previous section, it is still possible to compute the similarity between the refined observation vector and refined fingerprint vectors by applying a kernel trick, described in Section~\ref{ss:kernel}.

\subsection{Beacon Selection Strategy}
\label{ss:selection}
In light of the empirical insight obtained in Section~\ref{ss:locationEstimation}, we propose a novel beacon selection strategy to exclude the beacon that provides sensitive RSS measurements. By sensitive, we mean that the RSS values change drastically with a subtle or no change in distances. 
Given the $N$ number of beacons defined in set $B$, our goal is to select a subset of beacons $B_s^{(l_i)} \subseteq B$ such that the total variance is minimized.
Intuitively, when the variance of individual RSS distribution is small, it means that the particular beacon can provide a more robust RSS measurement that is less sensitive to the subtle change in distances.

Even though we argue that the beacon located farther away can provide less sensitive RSS measurements, it is not recommended to include very far away beacons for similarity computation. 
The signals from those very far away beacons can get lost frequently due to the undesirable propagation condition. 
Hence, we should also account for this factor in our beacon selection strategy. 
Combining the above two considerations, the objective function of our beacon selection strategy can be described as follows:
\begin{equation}
\begin{aligned}
B_s^{(l_i)} &= \argmin_{B_s^{(l_i)} \in B} \sum_{j = 1}^{s} \sigma^2(b_j ) \\
\text{s.t.} &\left|\frac{T_d}{T_a} - |\mathcal{R}_{ji}| \right| \geq \gamma, \forall b_j \in B_s^{(l_i)}
\end{aligned}
\label{eq:objFunc}
\end{equation}
where $\sigma^2(b_j)$ is the function computing the individual RSS variance from beacon $b_i$. 
Recall that we have a set of  instantaneous RSS values measured by a smartphone located at $l_i$ over a scanning duration $T_d$, i.e., $\mathcal{R}_{ji} = \{p(d_{ji}, t) | 0<t\leq T_d, b_j \in B \}$, then $\sigma^2(b_j)$ can be computed as follows:
\begin{equation}
\sigma^2(b_j) = \frac{\sum_{t\leq T_d}(p(d_{ji}, t) - f_j)}{|\mathcal{R}_{ji}|}
\end{equation}
where the $j$-th element of fingerprint vector, i.e., $f_j$, obtained by Eq.~(\ref{eq:ta}), also denotes the average RSS over all the instantaneous RSS values containing in $\mathcal{R}_{ji}$.
Note that $B_s^{(l_i)}$ might not necessary containing the same set of beacons as in $B_s^{(l_j)}$. 
In other words, those beacons that are not selected for similarity computation for fingerprint registered in $l_i$ might be selected for fingerprint registered in $l_j$. 
Each registered fingerprint has its own optimum set of selected beacons that is useful for similarity computation.

In Eq.~(\ref{eq:objFunc}), we also define a threshold $\gamma$ to check the number of received signals for all beacon $b_j \in B_s^{(l_i)}$. 
This $\gamma$ can be set according to the scanning duration of the smartphone and the advertising interval $T_a$ of the beacon. 
Consider the example where $T_a = 100$~ms, then the smartphone should expect no more than $10$ signals per second.
Hence, if the smartphone is configured to scan, say $T_d = 10$~s, then it should expect no more than $100$ signals. 
However, since the beacon starts broadcasting at a random time, it is very unlikely to receive  $10$ signals per second, even assuming an ideal channel condition. 
Suppose that the maximum signal lost (from expected) that we can tolerate is 20\%, then we can define $\gamma$ according to the following relation:
\begin{equation}
\gamma = \frac{T_d}{T_a} (1 - \eta)
\end{equation}
where $\eta$ is the maximum signal lost rate that we can tolerate.
In general, it is not recommended to have $\eta \leq 0.5$ because when the signal lost rate is more than half of the receiving rate, it simply indicates that the particular beacon is not the best choice to be selected. 
Imagine in a practical scenario where the smartphone always cannot receive the signal from this beacon, then this beacon does not have a significant impact on the similarity computation most of the time.

While our beacon selection strategy allows us to use a subset of less sensitive but more robust signals from selected beacons for similarity computation, it reduces the dimensionality of the observation vector and fingerprints to $s = |B_s^{(l_i)}|$. 
In Section~\ref{sec:ex}, we examine the various configurations of $s$ during the training and validation before deciding the best configuration for $s$.
Such a dimensionality reduction induces another problem to similarity computation, especially when the linearly separable vectors now become linearly non-separable. 
One solution is to map the vector in the current space to the high-dimensional space such that the vectors become linearly separable again.
However, it is computationally expansive to compute the similarity in the high-dimensional space, not to mention the additional mapping function to transform the vector from one space to another space.
To address this issue, this paper employs a kernel trick to compute the similarity between any two vectors in the high-dimensional space without literally visiting the space~\cite{scholkopf2001kernel}.

\subsection{Kernel Method for Similarity Computation}
\label{ss:kernel}
Suppose that we have a mapping $\phi : \mathbb{R}^N \rightarrow \mathbb{R}^H$ that transforms the fingerprint in the input space to some high-dimensional feature space $\mathbb{R}^H$. 
We can compute the similarity with a kernel function, $\kappa(\mathbf{f}_i, \mathbf{o}) = \phi(\mathbf{f}_i)^T\phi(\mathbf{o})$, which it computes the inner product of these vector in the $H$-space.
Mathematically, the similarity computation in the $H$-space can be defined as follows:
\begin{equation}
\begin{aligned}
\beta (\mathbf{f}_i, \mathbf{o}) &= \frac{\langle \phi(\mathbf{f}_i), \phi(\mathbf{o}) \rangle }{||\phi(\mathbf{f}_i)|| \, ||\phi(\mathbf{o}||} \\
& = \frac{\kappa(\mathbf{f}_i, \mathbf{o})}{\sqrt{\kappa(\mathbf{f}_i, \mathbf{f}_i) \kappa(\mathbf{o}, \mathbf{o})}} 
\end{aligned}
\label{eq:kernel}
\end{equation}
Let $\mathcal{D}$ be the distance metric, we can convert the distance metric, discussed in Section~\ref{ss:distMetrics}, to a kernel with the following two methods:
\begin{itemize}
	\item $\kappa(\mathbf{f}_i, \mathbf{o}) = \exp(-\mathcal{D} \gamma)$
	\item $\kappa(\mathbf{f}_i, \mathbf{o}) = \frac{1}{\mathcal{D}} \max(\mathcal{D})$
\end{itemize}
To perform the similarity computation in the input space, we have $\beta (\mathbf{f}_i, \mathbf{o}) = \mathcal{D}$. 
Based on the above two methods, we can develop a few kernel functions that satisfy Mercer's Theorem~\cite{steinwart2012mercer}, which states that the kernel should be a positive definite function for performing the inner product in the $H$-space.

Even though there are many kernel functions we can develop based on the above two methods, one important function development decision is to define a kernel function that fixes our problem context, while ensuring the developed kernel function satisfies Mercer's Theorem.
Before we can define the kernel function for our problem, we use the  Reproducing Kernel Hilbert Space (RKHS)~\cite{steinwart2012mercer} to examine the induced $H$-space.
RKHS can be constructed by considering the mapping $\phi : \mathbf{f}_i \rightarrow \kappa(\cdot, \mathbf{f}_i)$, which measures the similarity between the input vector with all the fingerprints in the database.
Consider the kernel function in the form of $f(\cdot) = \sum_{i =1}^{n} \mathbf{a}_i \kappa(\cdot, \mathbf{f}_i)$ for $n>0$, $\mathbf{a}_i \in \mathbb{R}^H$, and $\mathbf{f}_i \in \mathbb{R}^s$.
Here we refer to the refined $\mathbf{f}_i \in \mathbb{R}^s$ containing the RSS values from the set of selected beacons.
Then, we can establish the $H$-space by performing an inner product for these two elements, 1) $f(\cdot) = \sum_{i =1}^{n} \alpha_i \kappa(\cdot, \mathbf{f}_i)$ and 2) $g(\cdot) = \sum_{j =1}^{m} \mathbf{b}_j \kappa(\cdot, \mathbf{f}_j)$ as:
\begin{equation}
\langle f, g \rangle = \sum_{i =1}^{n} \sum_{j =1}^{m} \alpha_i \mathbf{b}_j \kappa(\mathbf{f}_j, \mathbf{f}_j)
\end{equation}
In particular, the $H$-space associated with a kernel $\kappa$ can be established by completing the norm $||f|| = \sqrt{\langle f, f\rangle}$.
Based on the inner product in the RKHS, we can refine the similarity computation based on kernel function to the following:
\begin{equation}
\begin{aligned}
\beta (\mathbf{f}_i, \mathbf{o}) &= \left\langle 
\frac{\kappa(\cdot, \mathbf{f}_i)}{\sqrt{ \langle \kappa(\cdot, \mathbf{f}_i), \kappa(\cdot, \mathbf{f}_i) \rangle }}, 
\frac{\kappa(\cdot, \mathbf{o})}{\sqrt{ \langle \kappa(\cdot, \mathbf{o}), \kappa(\cdot, \mathbf{o}) \rangle }} \right\rangle
\end{aligned}
\label{eq:kernelRKHS}
\end{equation}

When an arbitrary time is considered, the RSS variations measured by the smartphone are relatively small.
Hence, we can choose a stationary kernel such that $|| \phi(\mathbf{f}_i) ||^2 = \kappa(\mathbf{f}_i, \mathbf{f}_i) = \kappa(0)$.
Among all the possible kernels, the Gaussian kernel satisfies our above arguments since it maps all the data points on the same orthant, ensuring $0 \leq \beta (\mathbf{f}_i, \mathbf{o}) \leq 1$.
Considering a set of unique fingerprints $\mathbf{f}_1, \mathbf{f}_2, \cdots, \mathbf{f}_M$, the mapped data points $\phi(\mathbf{f}_1, \mathbf{f}_2, \cdots, \mathbf{f}_M)$ spans an $M$-dimensional subspace based on the infinite dimensional RKHS defined on the $H$-space.
By using the Gaussian kernel, we can compute the similarity between $\mathbf{f}_i$ and $\mathbf{o}$ in the $H$-space as follows:
\begin{equation}
\beta (\mathbf{f}_i, \mathbf{o}) = \exp(-\frac{|| \mathbf{f}_i - \mathbf{o} ||^2}{2\sigma^2})
\end{equation}
where $\sigma$ is a parameter that controls the width of the Kernel. Finding the best $\sigma$ is not the focus of this work. Instead, we tune the $\sigma$ during the training until optimum performance is observed. By optimum performance, we mean there is no more performance gain with further parameter tuning.



\section{Empirical Analysis}
\label{sec:ea}
Empirical analysis is conducted to investigate the noise present in the collected fingerprints, and also study the signal variations by comparing the fingerprint obtained through the offline stage and the fingerprint observed during the online stage.

\subsection{Outliers in the Offline Registered Fingerprints}
The common approach to compute Eq.~(\ref{eq:fj}) is by averaging the RSS data we collected during the on-site fingerprint survey.
Suppose that the beacon broadcasts its signal every time interval $T_a$ ($T_a$ is known as advertising interval in BLE terminology), then we should expect $1/T_a$ signals per second. 
During the offline stage, the receiver is configured to scan for a much longer duration so that we can collect sufficient data for fingerprint construction purposes.
Let $\mathcal{R}_{ji} = \{p(d_{ji}, t) | 0<t\leq T_d, b_j \in B \}$ be a set of instantaneous RSS values from beacon $b_j$, acquired by the receiver located at $l_i$ over a scanning duration $T_d$, then Eq.~(\ref{eq:fj}) can be computed as follows:
\begin{equation}
f_j^{(l_i)} = \frac{1}{|\mathcal{R}_{ji}|}\sum_{\mathcal{R}_{ji}}  p(d_{ji}, t)
\label{eq:ta}
\end{equation}
Rather than computing the time average on the raw RSS data, it is better to perform some data pre-processing to get rid of possible outliers.

\begin{figure}  
	\centering
	\includegraphics[width=1\columnwidth]{ 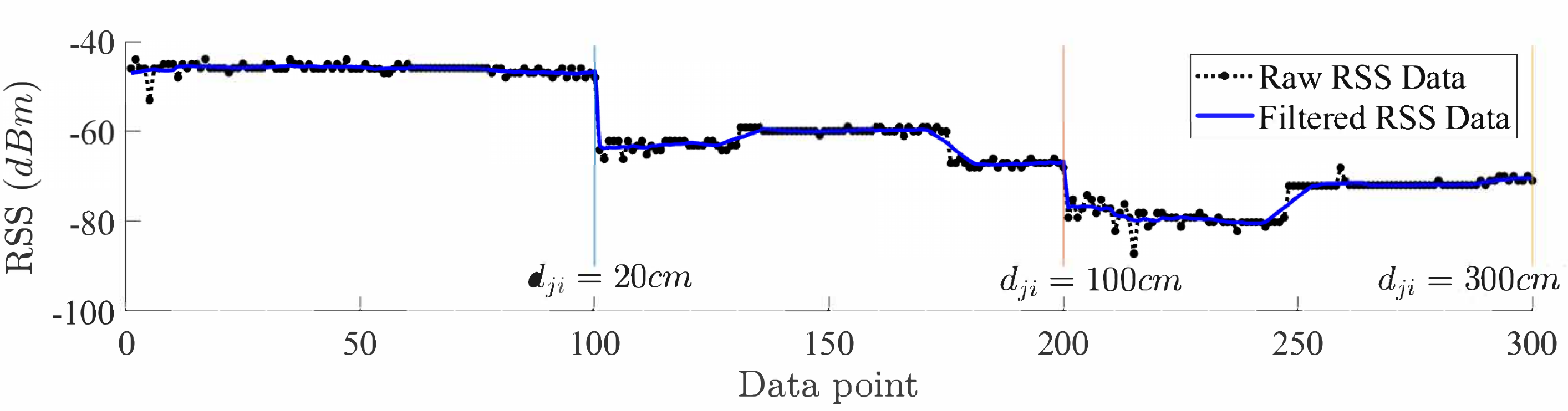}
	\vspace*{-0.5cm}
	\caption{After pre-processing, we can obtain a much smoother RSS data that gets rids of possible outliers. 
	}
	\label{fig:outlierAnalysis}
\end{figure}

The data we collected by fixing the receiver at a certain distance from the beacon is shown in Fig.~\ref{fig:outlierAnalysis}.
We plotted the RSS data obtained from 3 distances (i.e., $d_{ji} =$ 20~cm, 100~cm and 300~cm) for quick visualization.
The receiver was configured to scan for at least $10$~s, and the beacon was set to advertise at every $100$~ms.
Hence, there are at least $100$ data points from each distance. 
For a fair comparison, we only plotted the first 100 data points for each distance, as shown in Fig.~\ref{fig:outlierAnalysis}.
We can see that there are some unexpected outliers at each distance.
These random outliers simply indicate that it is not recommended to use the raw RSS data for fingerprint construction, rather some filtering methods can be applied to get rid of these outliers.

In this demonstration, we applied the moving average with a window size equal to ten. 
This number is chosen because this is a reasonable window size considering the number of signals broadcast by the beacon every second.
That is, if $T_a = 100$~ms, the receiver should observe no more than $10$ signals per second. 
Also, it is very rare for a user to move for more than $2$~m in one second.
From the filtered curve as well as the histograms shown in Fig.~\ref{fig:outlierAnalysis}, we can see that the variance of the data for each distance is reduced after applying the moving average.
Hence, it is recommended to pre-process the raw RSS data before computing the time average RSS.

Besides moving average, we can also consider kalman filter~\cite{8646581}, belief condensation filter~\cite{8683560} or particle filter~\cite{8994072}. 
In this paper, we opted to use the moving average for its simplicity in implementation and we also show that in Section~\ref{sec:ex}, the choice of filtering method does not have a significant impact on our proposed location estimation method as long as they can efficiently get rid of some outlier during the fingerprint construction.
All the fingerprints collected during the on-site survey will be stored in an online database.
The database should contain the information related to the location grid points for each fingerprint so that this information can be retrieved to estimate the location.

\subsection{Signals Variations during the Online Observation}
\label{ss:finding}
While height is not the major consideration for our work, we realized that the height of a smartphone is a major factor affecting the accuracy of location estimation. Recall Fig.~\ref{fig:estProblem}, it is clear that a subtle change in heights can return a very different RSS measurement even if the smartphone is in the same $x$ and $y$ coordinates. 
Motivated by the above argument, we consider the beacon selection a critical step prior to location estimation.
Intuitively, we need to select a subset of beacons that provide a more insensitive RSS value with respect to the change in heights.
The change in RSS values should be insignificant when the change in distances (i.e, the distance between the smartphone and the beacon) is less than 1~m.
This is because different users might carry the smartphone at different heights from the floor.
However, this distance is generally less than 1~m if we assumed the average maximum height of people is about 190~cm.

\begin{figure}
	\centering
	\includegraphics[width=0.85\columnwidth]{ 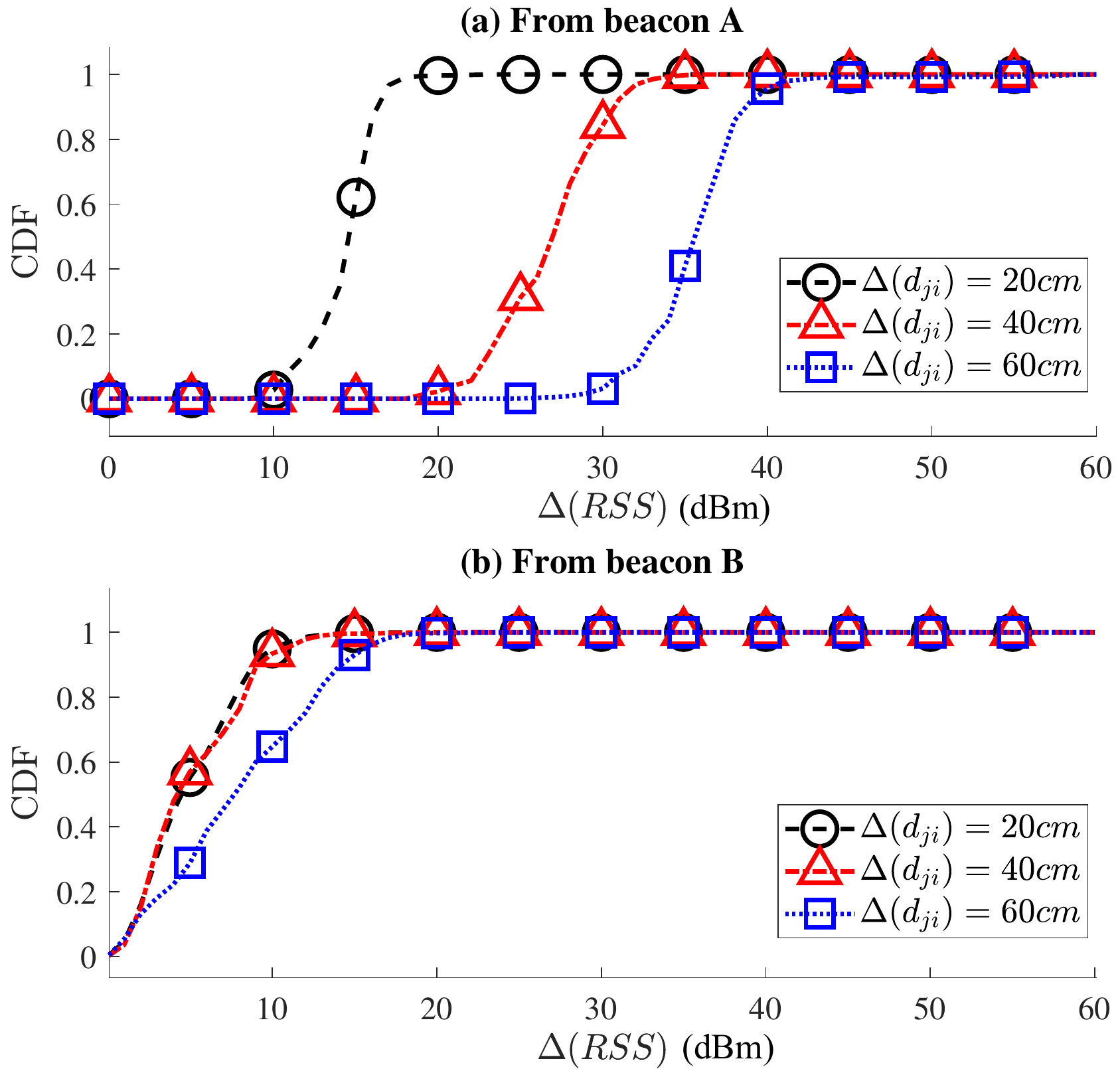}
	\caption{Beacon A was deployed at a location very close to the grid point $l_j$, whereas beacon B was located 4~m away from $l_j$. The CDF plots indicate RSS variations for three different changes in distances. The change in the distances indicates the holding dynamic of a smartphone while the user remains still in the same grid point.}
	\label{fig:variationDistAnalysis}
\end{figure}
An experiment is conducted to verify the change of RSS values ($\Delta(RSS) = RSS_{\text{ref}} - RSS_{\text{measured}}$) with respect to a slight change in distances (i.e., less than 1~m). 
The $RSS_{\text{ref}}$ is the reference value from the registered fingerprint obtained during the offline survey.
A user was asked to hold a smartphone freely while staying still in the same grid point $l_j$.
The experiment consists of two beacons: beacon A was located very near to the user, while beacon B was located at 4~m away from the user.
The CDF plots for the change in RSS values with respect to the change in distances are shown in Fig.~\ref{fig:variationDistAnalysis}.
From the plots, we can observe a drastic change in RSS values from beacon A  when the user only moved the smartphone for about 20~cm from its initial holding position, i.e., $\Delta(RSS)$ is about 16~dBm for $80\%$ of the time.
When the user moved the smartphone for about 40~cm and 60~cm, $\Delta(RSS)$ can go up to 30~dBm and 38~dBm for $80\%$ of the time, respectively.
However, the RSS values from beacon B show fairly robust values despite the small change in distances.
When the distance is less than or equal to 40~cm, $\Delta(RSS)$ is generally less than 10~dBm for more than $90\%$ of the time.
When the user moved the smartphone 60~cm from its initial position, $\Delta(RSS)$ is still less than 15~dBm for more than $90\%$ of the time.

Many works claim that the stronger the RSS value, the better the estimation~\cite{abed2019rss, 6846747, ezhumalai2021efficient}. 
Our empirical analysis, surprisingly, unveils that those beacons which contribute the strongest RSS might adversely affect the estimation performance. 
In light of the above discovery, we propose a novel beacon selection strategy that refines the online observation prior to location estimation.
Note that some works in ToA estimation also show that the strongest path is not the favorable path for the location estimation~\cite{1570024, 6062690}. These works mostly argue from the context of multipath propagation. Since the signals arrived from multiple paths can be added constructively or destructively, the strongest signal might not indicate the direct path. While the back-search approach is applied by these works to find the direct path~\cite{6062690}, such a back-search approach cannot be directly applied to our problem scenario considering the multiple signals sources from different beacons at different locations. 
Even though it is possible to identify the direct path from all the beacons, the signals from all the direct paths of all the signals are still sensitive to the small change in the holding position of a smartphone, which is unfavorable to our problem scenario. 
Rather than detecting the direct path, our work aims to select the signal sources that produce a more robust RSS measurement for fingerprinting.

This section describes the setting of our experimental testbed for fingerprint data collection.
The collected data is consolidated into training and testing datasets and is made publicly available in Github\footnote{RSSI BLE Localization UoG, "https://github.com/pspachos/RSSI-BLE-Localization-UoG"} and IEEE DataPort\footnote{RSS FINGERPRINT DATA, "https://ieee-dataport.org/documents/rss-fingerprint-data"}.

\subsection{Experimental Testbed}
\begin{figure}
	\centering
	\includegraphics[width=1\columnwidth]{ 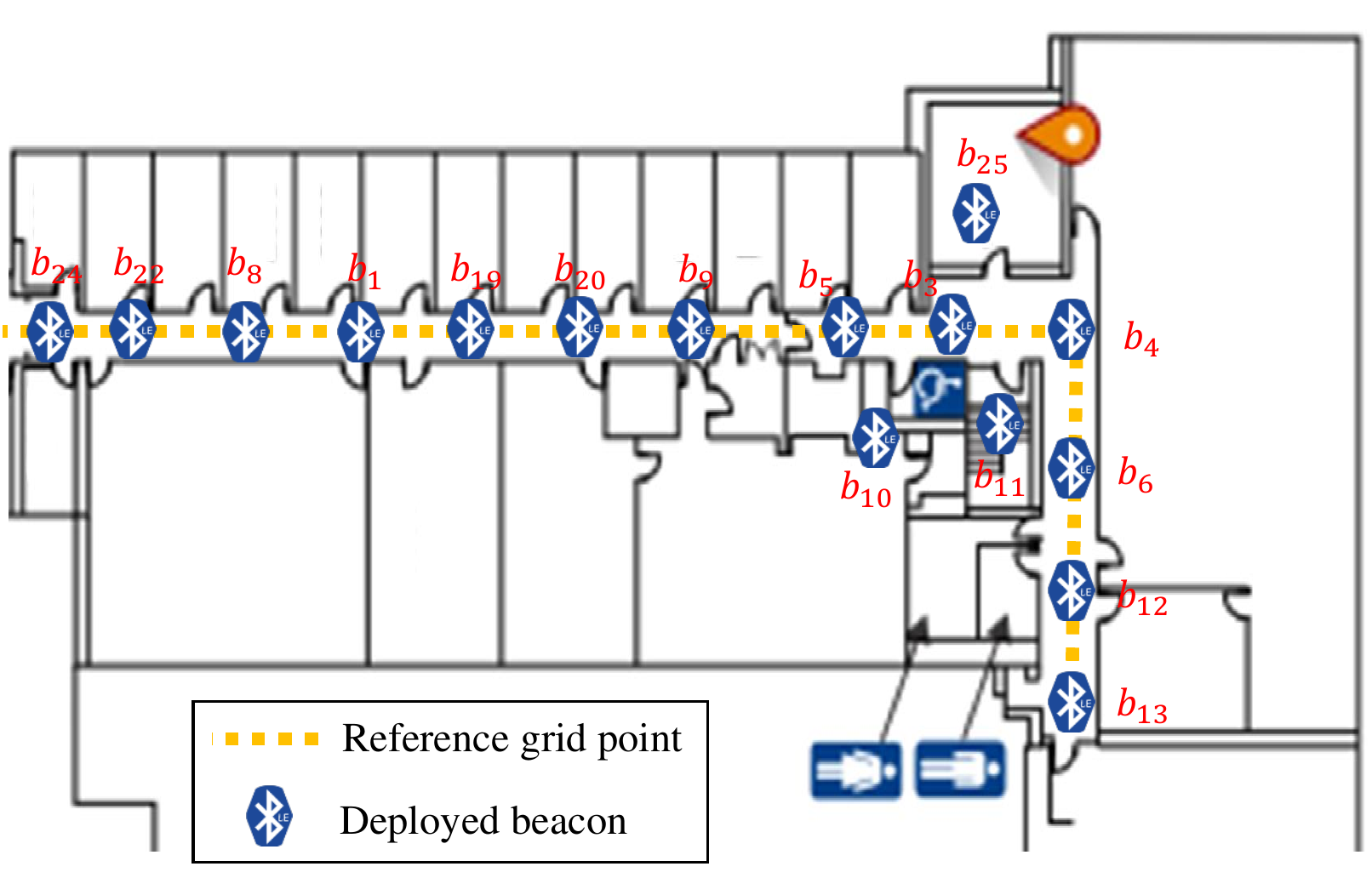}
	\vspace*{-0.5cm}
	\caption{The floorplan of the experimental testbed.}
	\label{fig:floorplan}
\end{figure}

\begin{figure}
	\centering
	\includegraphics[width=.85\columnwidth]{ 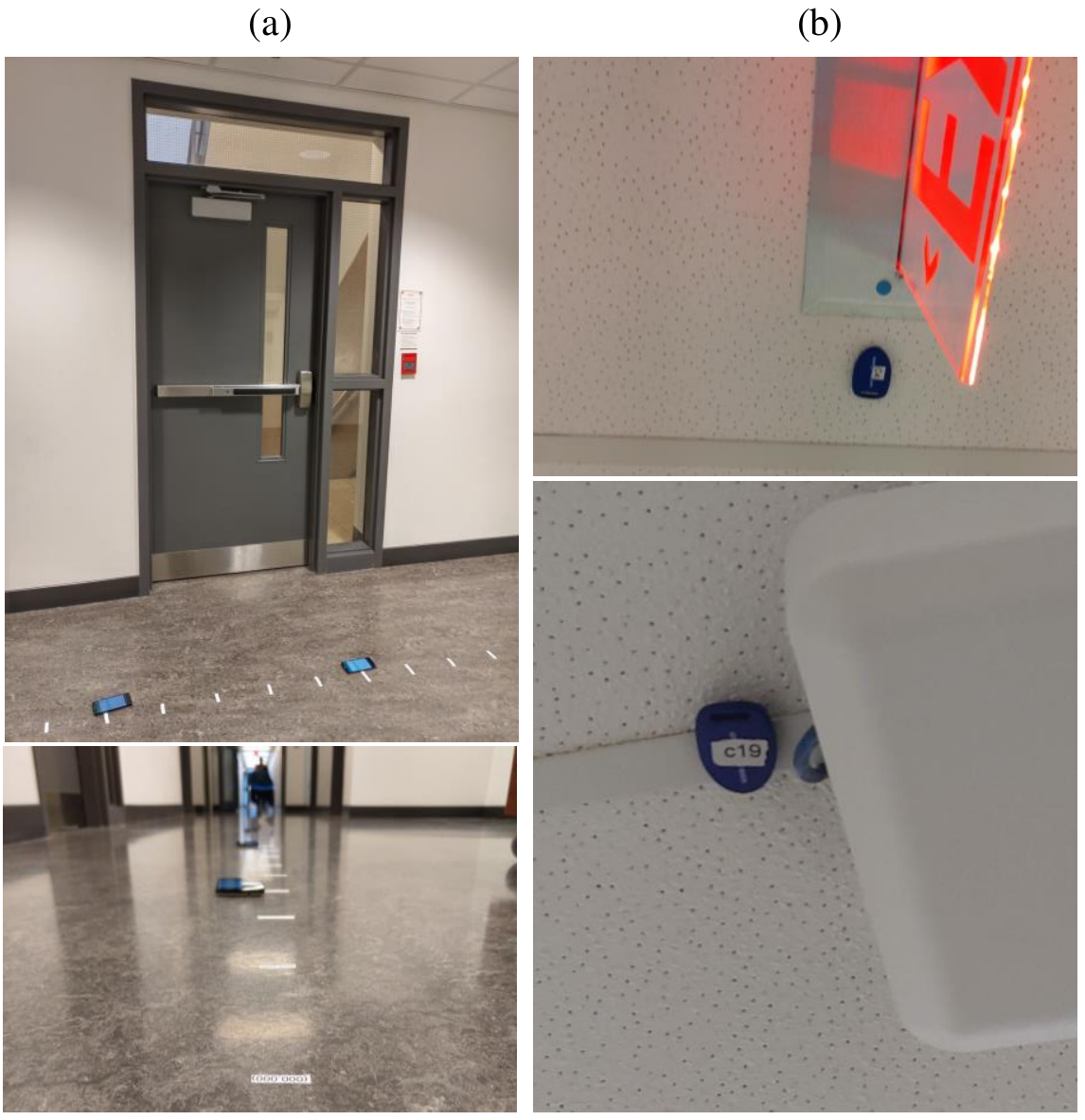}
	\caption{(a) The smartphones were placed on each grid point for the on-site fingerprint survey. (b) The beacons were deployed on the ceiling.}
	\label{fig:deployment}
\end{figure}

The floorplan of our experimental testbed is shown in Fig.~\ref{fig:floorplan}.
The yellow dotted lines refer to the reference grid points, whereas the Bluetooth symbols indicate the deployed beacons.
A few smartphones were programmed to work as a scanner for fingerprint surveying purposes.
The on-site fingerprint survey with our preprogrammed smartphones is illustrated in Fig.~\ref{fig:deployment}(a).
There were a total of 16 beacons installed on the ceiling, as shown in Fig.~\ref{fig:deployment}(b).
Some beacons were deployed at the location very near to the WiFi access point and the exit signboard.
Even though we can have a better deployment plan to avoid such undesirable locations, we purposely deployed a few beacons at these locations to show that our proposed approach can still produce good performance under the influence of WiFi signals and other obstacles.

\section{Experimental Data Collection}
\label{sec:ex}

\begin{figure}
	\centering
	\includegraphics[width=1\columnwidth]{ 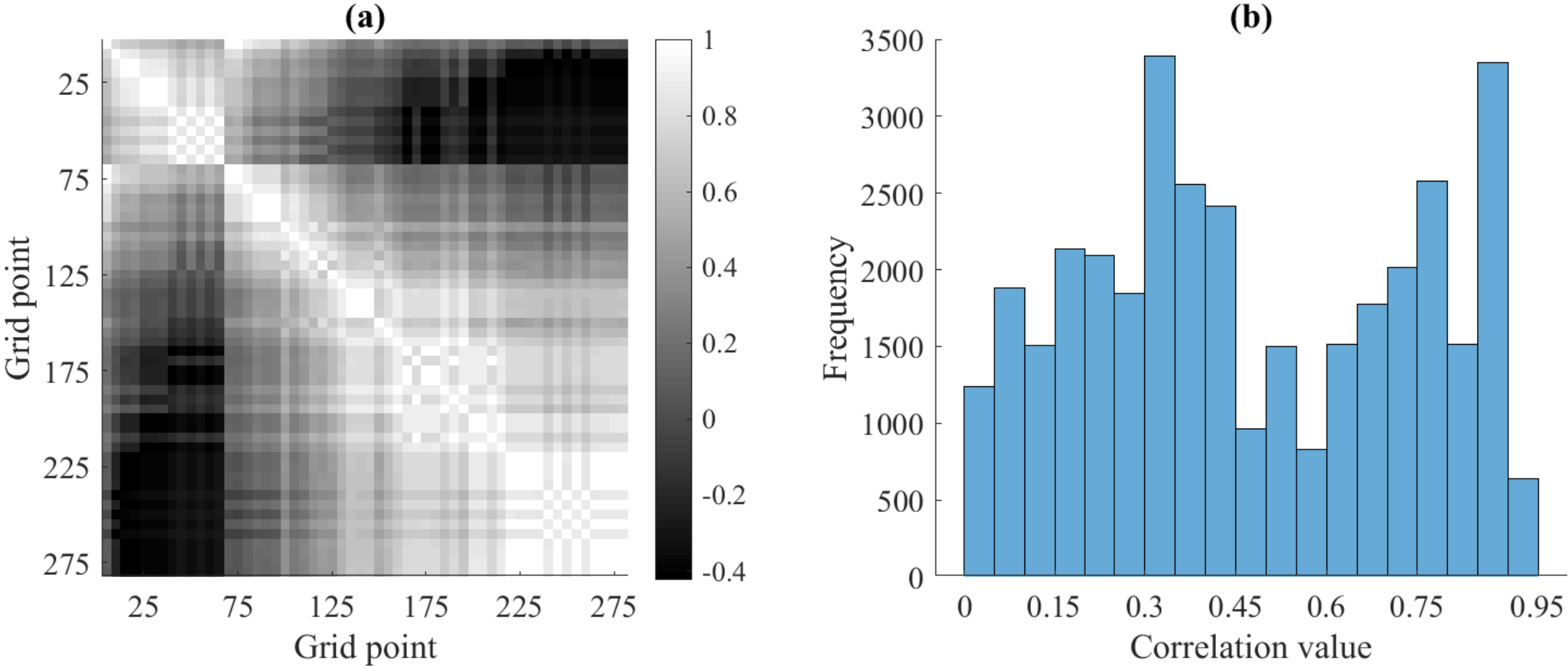}
		\vspace*{-0.3cm}
	\caption{(a) The correlation score between each fingerprint at the corresponding grid point. (b) The histogram of the correlation score.}
	\label{fig:corrFingerMap}
\end{figure}
\subsection{Fingerprint Construction}
A total of 276 grid points were labeled according to their $x$ and $y$ coordinates. 
The smartphone was configured to scan for 30~s at each grid point during the on-site fingerprint survey. 
Since each beacon was configured to broadcast the signal every 100~ms, the smartphone shall expect at most 10 $\times$ 30 signals from each beacon at each grid point during the on-site fingerprint survey.
The reason to have a long scanning duration during an on-site fingerprint survey is to obtain sufficient signals for constructing the fingerprint database.
Furthermore, we can have a more robust fingerprint representation by averaging RSS values from the same beacon rather than using a single raw value.

According to Definition~\ref{def:LF}, each fingerprint is a unique representation, in which the similarity between each fingerprint should be less than one.
The correlation between each fingerprint is computed and depicted in Fig.~\ref{fig:corrFingerMap}(a).
The correlation score is always 1 when computing the correlation of the fingerprint to itself.
Note that the heatmap does not provide a clear distinction for fingerprints at those closely spaced grid points because the correlation score of these closely spaced fingerprints is very close to one, but not one.
For further verification, we plotted the histogram by extracting the upper triangle of the correlation score matrix and excluding the diagonal elements.
The histogram, as shown in Fig.~\ref{fig:corrFingerMap}(b), confirms that the correlation score between all the other fingerprints is always less than one.

\begin{figure}
	\centering
	\includegraphics[width=1\columnwidth]{ 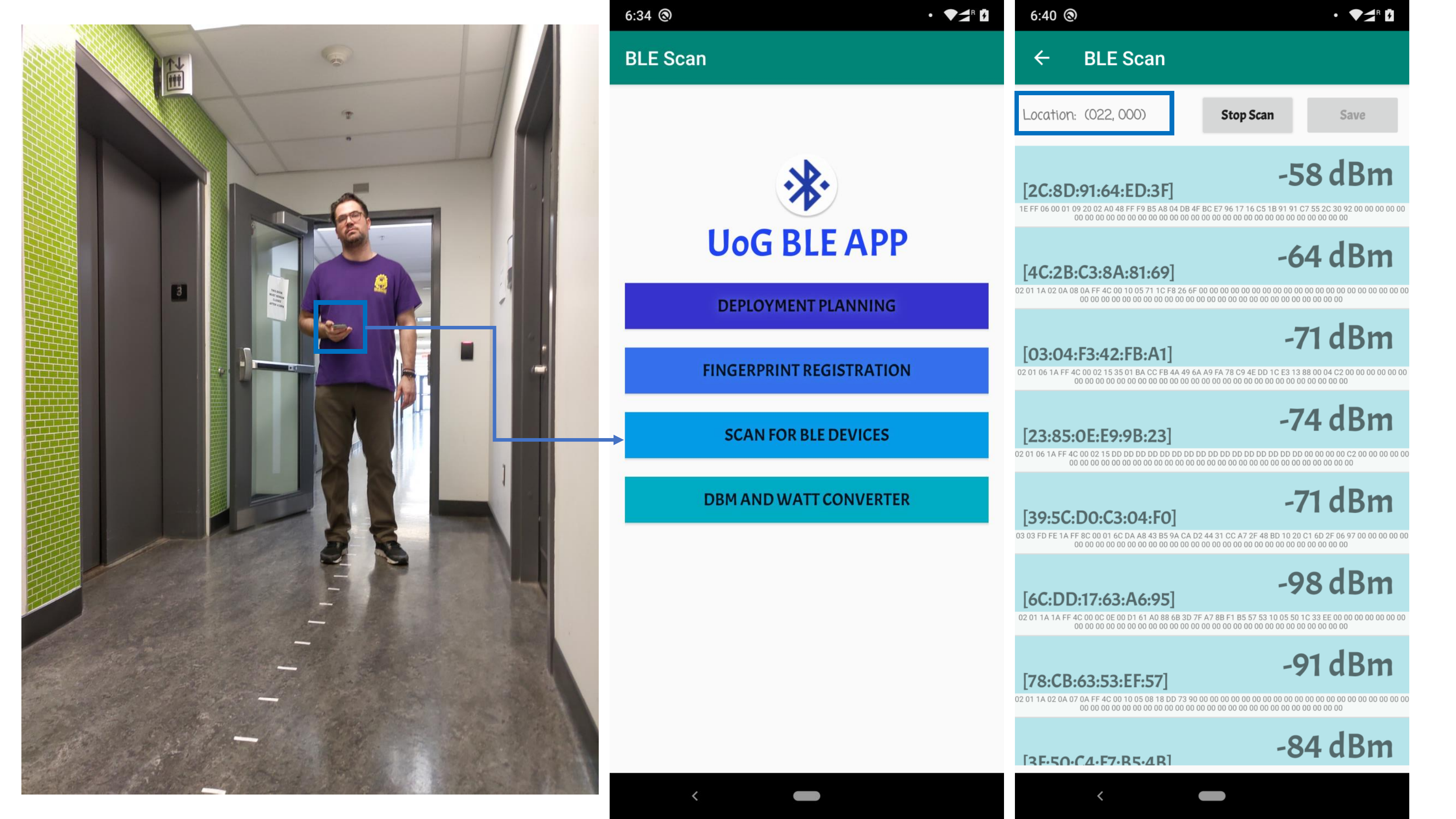}
		\vspace*{-0.5cm}
	\caption{A user was asked to hold the smartphone in his hand randomly while taking the testing data. While displaying the current location, we also configured the application to log all the data into the local storage for further experiments.}
	\label{fig:testingCollection}
\end{figure}

\subsection{Testing Data}
We can use all the 1027,038 raw RSS data collected during the on-site survey or the 276 constructed fingerprints for training purposes.
Regardless, the final fingerprint map is a matrix of dimension $25 \times 276$, where 25 indicates the size of the fingerprint vectors and 276 denotes the total number of grid points.
For the testing data, we used the data collected from: 1) placing the smartphone that was placed on the floor (\textit{type 1 data}), and 2) the smartphone that was held in the hand (\textit{type 2 data}). 
The purpose to collect \textit{type 1 data} is that we would like to verify the performance of similarity computation when the smartphone was placed at almost the same location as the one used during the on-site survey. 
This should give us a better result since the RSS variations with \textit{type 1 data} are mainly due to environmental factors, such as multipath and shadowing effects, rather than dynamic hand movements.
Then, we use this result to compare with the result obtained through \textit{type 2 data}, in which the RSS variations are further subject to dynamic hand movements.

For \textit{type 2 data}, two human subjects were asked to hold the smartphone in their hands freely while walking along a certain path following the grid point.
The height of these two human subjects is different, one is taller with a height approximately equal to 190~cm, and another with a height approximately equal to 165~cm.
Our experimental setup to collect the testing data from a human subject is illustrated in Fig.~\ref{fig:testingCollection}.
Both human subjects were assigned to follow different walking paths while starting the data collection simultaneously. Each time they would only stay at the grid point for less than 30~s to collect the data.
During the data measurement, the smartphone logged all the signals into the local storage for later experiments.
The following information was logged: the label of the grid point in $x$-$y$ coordinate, the RSS value, the signal's arrival time, and the beacon (i.e., the source of the signal).

The collected data is then exported as .csv files and imported to Matlab for further experiment.
Two protocols were designed to consolidate the raw RSS data into a list of observed fingerprints.
\begin{itemize}
	\item Protocol 1: the data was consolidated based on the prior knowledge of deployed beacons so that the length of the observed vector equals the number of deployed beacons.
	This consolidation method is impractical for real-time applications because it requires the smartphone to scan continuously until all beacons are seen.
	The data consolidated according to protocol 1 is mostly used to verify the location estimation performance in an ideal case. Then, we can compare the result to the practical case when the smartphone is unable to observe the signals from all beacons as a consequence of the short scanning duration.
	\item Protocol 2: the signal's arrival time is used to divide the data into 1~s duration. Specifically, the elapsed time is computed by taking the difference between subsequent arrival. 
	Based on the elapsed time, the signals can be segmented into 1~s each.
	Even though each beacon is configured to broadcast at every 100~ms, it is almost impossible to receive exactly 10 signals from each beacon due to unpredictable channel fluctuation.
	In other words, we may observe 10 signals in 1 s from beacon A, but only 2 signals in 1 s from beacon B. 
	Also, there might be a possibility that no signals are observed from some beacons. Hence, the length of the observed vector at a certain time might be smaller than the length of the fingerprint stored in the database. 
\end{itemize}
The total number of raw RSS data collected during testing is 371,595, while the total number of testing data for both types, after consolidation, is shown in Table~\ref{table:data}.

\begin{table}
	\caption{Total number of consolidated data}
	\label{table:data}
	\centering
	\begin{tabular}{p{3.5cm}ll}
		\toprule
		\cmidrule{1-3}
		Consolidation Method    &  Type 1 data & Type 2 data  					\\
		                        & (on the floor)   & (in the hand) \\
		\midrule
		(Protocol 1) scan until all beacons are seen 	 & 6917          & 3129    \\  \\
		(Protocol 2) based on 1~s scanning duration                 & 22977         & 5573 \\ 
		\bottomrule
	\end{tabular}
\end{table}

\begin{figure}
	\centering
	\includegraphics[width=.95\columnwidth]{ 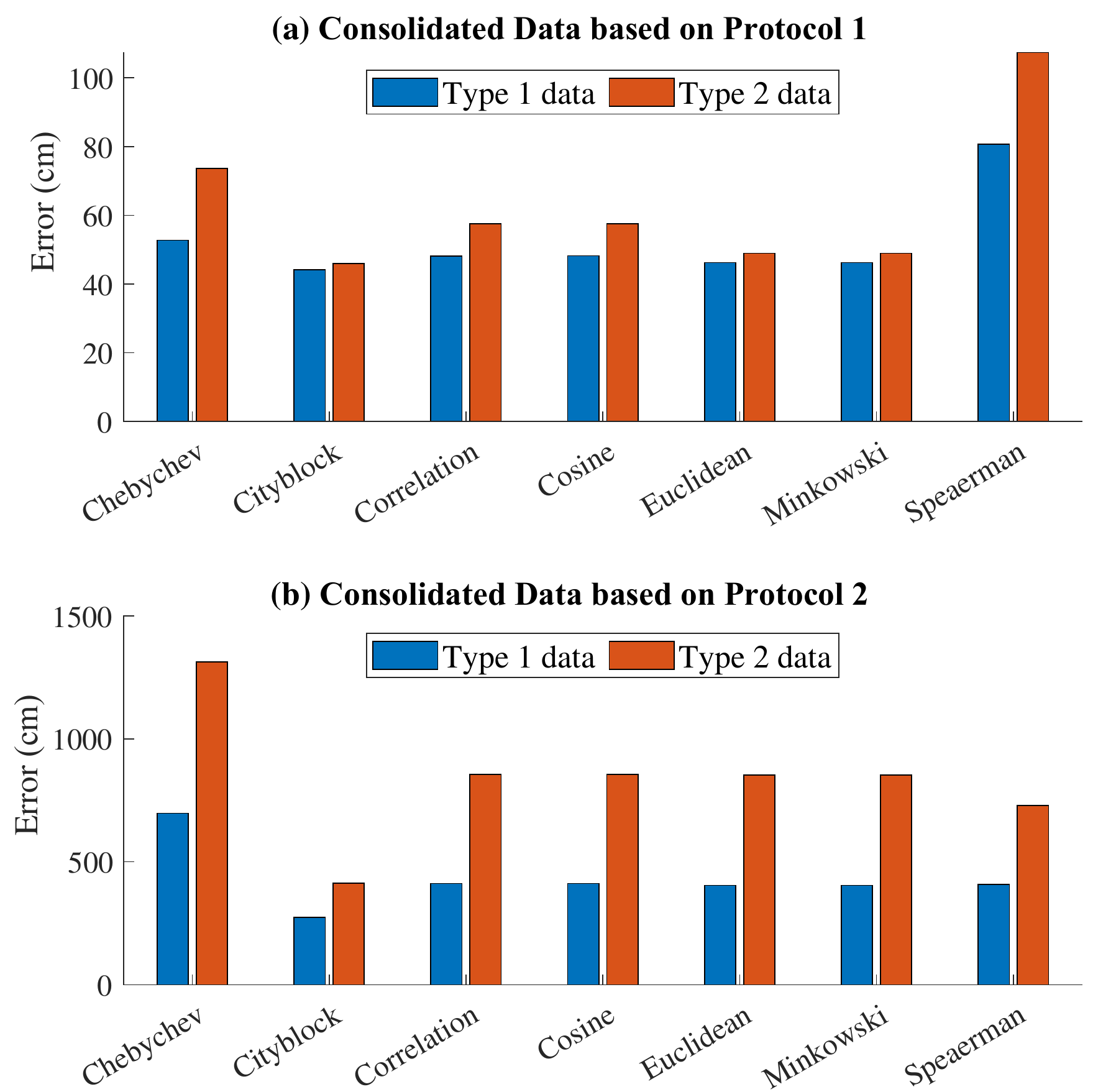}
	\caption{The localization error (in cm) produced by different similarity computations, for both types of data that are consolidated based on (a) protocol 1 and (b) protocol 2.}
	\label{fig:compareDistMetrics}
\end{figure}
\section{Results and Discussion}
\label{sec:ex2}
This section evaluates the localization performance based on the two types of data described above. First, we investigate the similarity computation with different distance metrics before selecting the best distance metric for performance comparison. Second, we use a validation approach to identify the optimum number of beacons for our beacon selection strategy. Lastly, we present an extensive performance evaluation comparing our proposed approach with the selected baselines and discuss the results.

\subsection{Baseline Selection}
We computed the average localization error achieved by the similarity computation with different distance metrics, including Chebyshev, Correlation, Cosine, Euclidean, Minkowski, and Spearman.
Fig.~\ref{fig:compareDistMetrics}(a) and (b) show the results for both types of data consolidated based on protocol 1 and protocol 2, respectively.
Note that type 1 data and the dataset consolidated based on protocol 1 are used mainly for performance validation.
In practice, this type of data is too ideal and impractical.
Rather, we are more concerned with the localization performance with type 2 data consolidated based on protocol 2, which reflects a more practical scenario where the data is observed during a 1~s scanning duration with a smartphone held freely in the hand.

As shown in Fig.~\ref{fig:compareDistMetrics}(a),  when protocol 1 is used to consolidate the observed fingerprint, the localization error is generally low.
Specifically, the error is no more than 1~m for all the distance metrics, regardless of data types.
For example, the localization error produced by Cityblock is 44.16~cm and 45.94~cm, for type 1 data and type 2 data, respectively.
That is, the difference between the two distance metrics is only 1.78~cm.
Such a small difference indicates that even though the smartphone might be held at different heights from the initial position where the fingerprint is surveyed, it is still possible to achieve satisfactory localization performance when the smartphone can observe sufficient signals from the same set of beacons.

Comparing Fig.~\ref{fig:compareDistMetrics}(a) and (b), we can see that the localization error is high when the data is consolidated based on the 1~s scanning duration.
The localization performance degrades when the smartphone fails to obtain sufficient observation for location estimation.
This is a common and practical problem when the smartphone is only configured to scan for a 1~s scanning duration.
Regardless of protocols, we can see that type 1 data generally produce lower error than type 2 data.
This verifies the challenge of location estimation when the smartphone is held in hand, compared to placing the smartphone at the exact grid point where the on-site fingerprint survey is conducted.
As discussed, type 1 data and the dataset consolidated based on protocol 1 are used to validate the performance in an ideal case.
On the other hand, our main goal is to improve the localization performance considering a practical scenario defined by type 2 data consolidated based on protocol 2.
Based on the results shown in Fig.~\ref{fig:compareDistMetrics}(a) and (b), we selected three distance metrics (i.e., Kernel, Cosine and Euclidean) as a baseline for performance comparison with our proposed method.

\begin{figure}
	\centering
	\includegraphics[width=1\columnwidth]{ 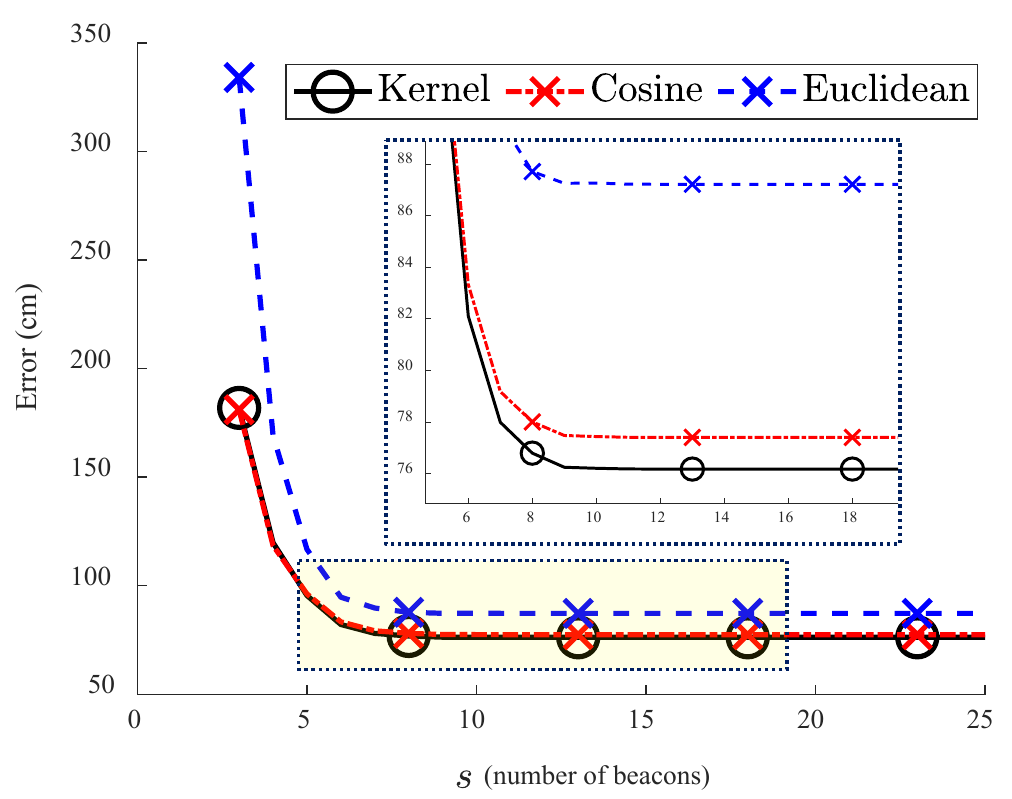}
\vspace*{-0.5cm}
	\caption{The error converges to a certain value when $s$ increases.}
	\label{fig:selectionValidation}
\end{figure}
\begin{figure*}
	\centering
	\includegraphics[width=1\textwidth]{ 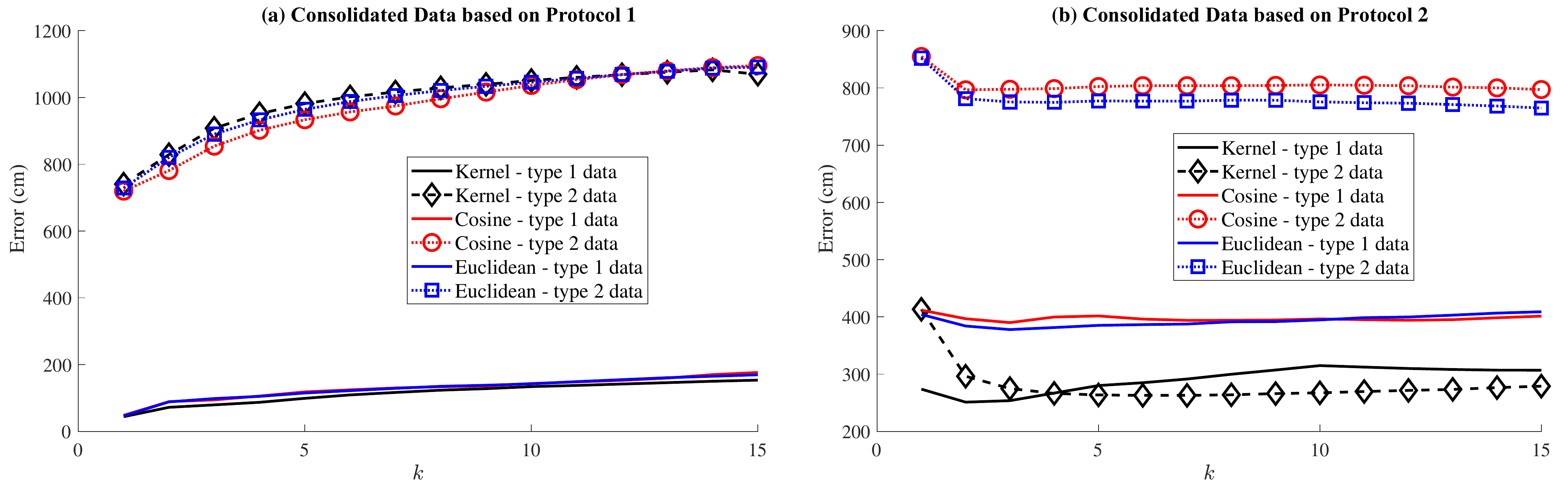}
	\vspace*{-0.5cm}
	\caption{The localization error (in cm) with respect to the number of neighbors, in which each neighbor contributes equal weight.}
	\label{fig:kNNResults}
\end{figure*}
\begin{figure*}
	\centering
	\includegraphics[width=1\textwidth]{ 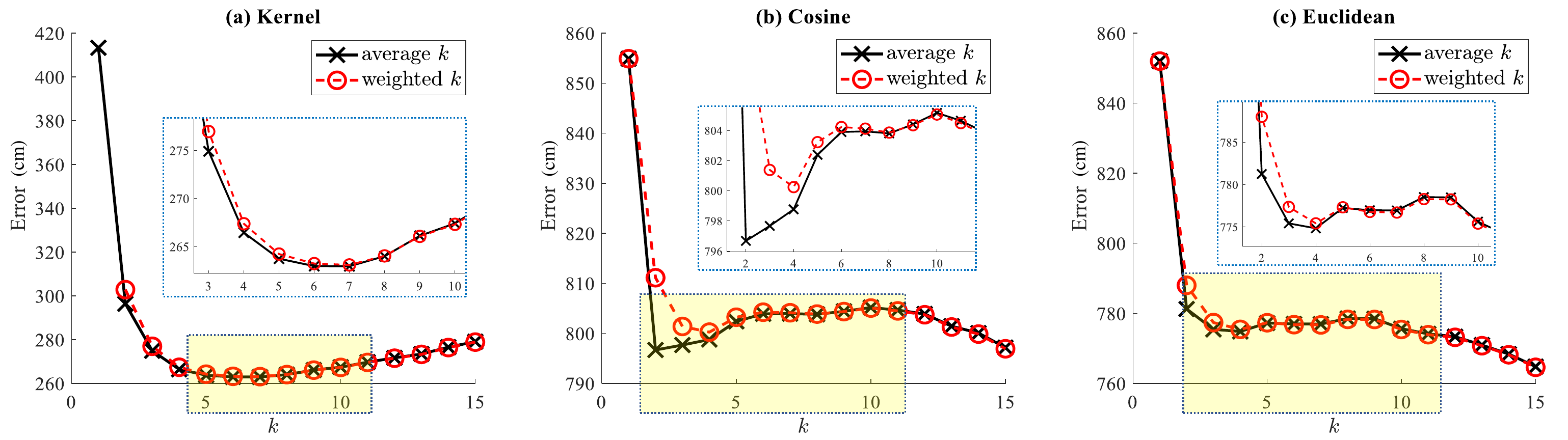}
	\vspace*{-0.5cm}
	\caption{The localization error (in cm) produced by (a) Kernel, (b) Cosine, and (c) Euclidean methods, considering the weight from top-$k$ neighbors.}
	\label{fig:weightedResults}
\end{figure*}
\subsection{Validating the Number of Selected Beacons}
Before proceeding to performance evaluation, we used the training data consisting of the raw RSS values collected during the on-site survey to validate the number of selected beacons, i.e., the parameter $s$ discussed in Section~\ref{ss:selection}.
Recall that our objective is to select a subset of beacons that have the minimum total variance, at the same time these beacons should provide sufficient signals for location estimation purposes.
While we can compute the total variance and estimate the number of signals directly to identify the subset of selected beacons, the underlying question is how many beacons we need to select such that to minimize the localization error.
Let $\mathcal{C}$ be the cost function for training, we have
\begin{equation}
\mathcal{C} = \frac{1}{n} \sum_{i = 1}^{n} \left(\mathbf{l}_i^{(T)} - \rho(\mathbf{f}_i^{(T)}) \right)^2 
\end{equation}
where $\mathbf{l}_i^{(T)} \in \mathbb{R}^2$ is a 2-dimensional vector representing the coordinate of the location grid point, the superscript $T$ indicates a training data. The function $\rho(\cdot)$ returns the estimated location given the fingerprint vector $\mathbf{f}_i$. 
Recall Eq.~(\ref{eq:estLoc}) and Eq.~(\ref{eq:setL}), we can define the function $\rho(\cdot)$ as follows:
\begin{equation}
\begin{aligned}
\rho(\mathbf{f}_i^{(T)}) &= \mathbf{w}^T \mathbb{L} \\
&= \sum_{i = 1}^{k} w_i \mathbf{l}_i \\
&= \sum_{i = 1}^{k} w_i \min_{L, k} \beta(\mathbf{f}_i, \mathbf{f}_i^{(T)})
\end{aligned}
\end{equation}
We can relate the above equation with $s$ based on the size of the fingerprint vector $\mathbf{f}_i$. That is, if we use all the beacons to define the fingerprint vector, then the dimension of $\mathbf{f}_i$ is $N$. If we only use a subset of selected beacons, then the dimension of $\mathbf{f}_i$ is $s$. 
Hence, the goal of our training is to find the number of selected beacons $s$ satisfying Eq.~(\ref{eq:objFunc}), while minimizing the cost function. 

The localization errors achieved by the similarity computation based on Kernel, Cosine, and Euclidean are shown in Fig.~\ref{fig:selectionValidation}.
We can see that the error decreases when $s$ increases and converges after a certain $s$.
When $s$ is greater than 5, the localization error is less than 1~m, for all three similarity computation approaches.
In other words, the localization performance is significantly improved reaching sub-meter accuracy when $s \geq 5$.
Based on this result, we chose to use $s=10$ for the rest of the experiment.

\begin{figure*}
	\centering
	\includegraphics[width=1\textwidth]{ 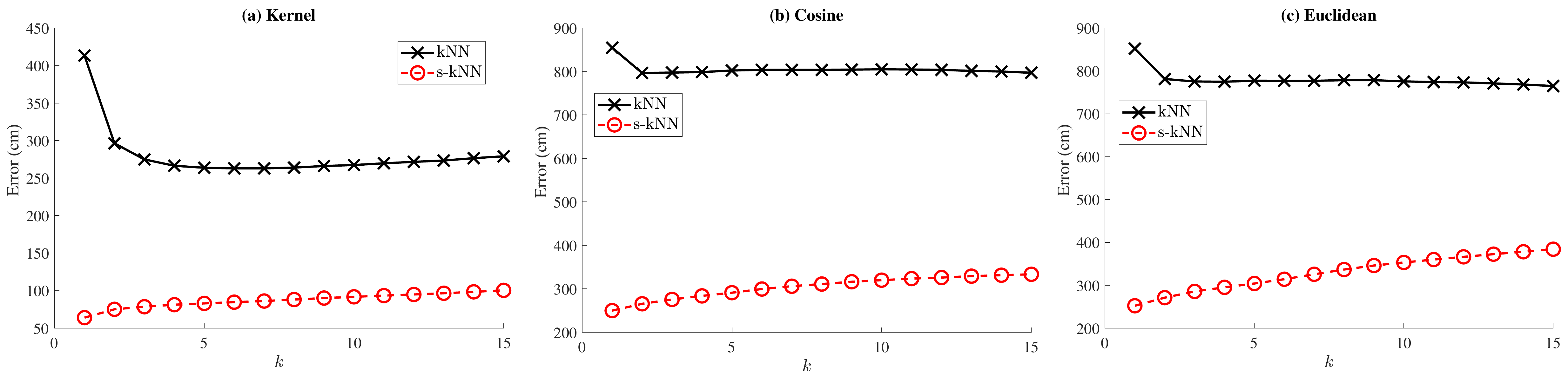}
		\vspace*{-0.5cm}
	\caption{The effect of beacon selection on the localization error (in cm) comparing conventional kNN and the kNN with our selection strategy (denote as s-kNN). Similarly, three metrics are used to compute the top-k similar neighbors, these distance metrics are (a) Kernel, (b) Cosine, and (c) Euclidean methods.}
	\label{fig:selectedResults}
\end{figure*}
\subsection{Performance Evaluation}
We further examined the effects of top-$k$ neighbors on the localization performance.
In particular, we estimate the location by averaging the top-$k$ reference grid points according to Eq.~(\ref{eq:estLoc}), i.e., the weights $\mathbf{w}$ are the same for all the $k$ grid points.
Recall that type 1 data is obtained with the smartphone placed on the floor, this type of data serves as the baseline for comparison with the dataset collected with the smartphone held by human hands (i.e., type 2 data).
The  localization errors with respect to the number of neighbors are shown in Fig.~\ref{fig:kNNResults}.
Comparing Fig.~\ref{fig:kNNResults}(a) and (b), we can see that the localization error with type 2 data is higher compared to type 1 data, when the data is consolidated based on protocol 1.
This can be explained by the nature of type 1 data, of which the data is collected at the exact same grid point where the fingerprint is surveyed.
In this case, the smartphone is capable of estimating the location with fewer errors when the smartphone can observe the exact same number of beacons that were used to survey the fingerprint.
However, when type 2 data is considered, the localization error is high even though the smartphone can observe the exact same number of beacons.
This verified our argument that not all the beacons contribute meaningful signals for location estimation.
While we cannot determine the holding dynamics of the smartphone, we can select the beacons that contribute useful signals for location estimation.

When the size of the observation is less than the number of beacons that were used to define the fingerprint during the on-site survey (consolidated data based on protocol 2), the kernel method performs better than the other two methods for both types of data, as shown in Fig.~\ref{fig:kNNResults}(b).
This again verified the advantages of using the kernel method for similarity computation when the size of the observation vector is different from those fingerprints registered in the database.
Fig.~\ref{fig:kNNResults} (b) shows that the kernel method can even achieve a comparable performance for both types of data.
Such a result is encouraging as it confirms that it is possible to achieve good localization performance even though the smartphone is held freely on the hand.

Previously, we considered the equal contribution from all the top-$k$ neighbors while estimating the location.
Such an averaging method may result in a great error especially when the other neighbors have the least impact on the location estimation.
Instead of considering the equal contribution, we used a weighting approach to accounting for the contribution from the top-$k$ neighbors.
In particular, we used the similarity score produced by each neighbor to define the weight.
Our experimental results show that there are only slight differences from the averaging approach shown in Fig.~\ref{fig:kNNResults}(a) and (b).
Hence, instead of presenting similar plots as in Fig.~\ref{fig:kNNResults}(a) and (b), we highlighted the differences by plotting the localization error of each similarity computation method.
Note that only type 2 data consolidated based on protocol 2 is used for this purpose.
The results achieved by (a) Kernel, (b) Cosine, and (c) Euclidean methods are shown in Fig.~\ref{fig:weightedResults}.
Interestingly, we see that the weighting approach does not achieve much performance gain compared to the averaging approach.
This is because the similarity of these top-$k$ neighbors are in fact very close and thus their contribution to the location estimation is almost equal.

In general, the localization error decreases when $k$ increases.
However, for Kernel and Cosine methods, we can see that the localization error increases again after $k$ reaches a certain value.
This is completely opposite to the Euclidean method, in which the localization error continues to decrease when $k$ increases.
This is mostly due to the different neighbors identified by all these three similarity computation methods.
In most cases, it is observed that the localization performance is good when $k$ is between 2 to 10.
For example, the Kernel method shows only 2.63~m error when $k=7$, the Cosine method achieves 7.97~m error when $k=2$, whereas the Euclidean method, achieves 7.74~m error when $k=4$.
Since our proposed selection approach suggested selecting the signals from at least 10 beacons for location estimation, then $k$ should not be greater than 10.

\begin{figure*}
	\centering
	\includegraphics[width=1\textwidth]{ 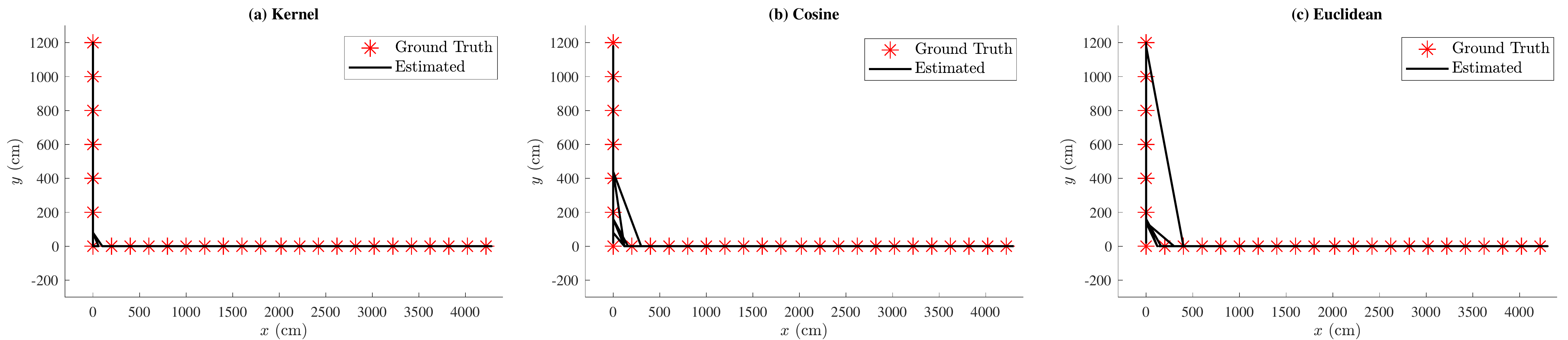}
		\vspace*{-0.5cm}
	\caption{The red markers indicate the walking path, whereas the black line depicted the estimation output produced by (a) Kernel, (b) Cosine, and (c) Euclidean.}
	\label{fig:walkpath}
\end{figure*}

\subsection{Performance Comparison}
So far, the kernel method achieves the best localization performance with accuracy up to 2~m when $k\geq 3$.
Next, we compared the result achieved by our proposed selection strategy to the above kNN that relies merely on the top-$k$ neighbors for performance improvement.
Note that kNN is the common approach used by many fingerprint-based location estimations, for example, \cite{7103024} used kNN to estimate the location based on the fingerprint constructed from BLE beacons, whereas \cite{7307751} used kNN with WiFi-based fingerprint for location estimation.
However, conventional kNN matches the fingerprint directly by comparing the observation vector and the fingerprint database without considering any selection strategy. 
To verify that our selection strategy can further improve the performance of kNN, we evaluated the effect of the number of neighbors on the localization performance for the fingerprint refined by our selection strategy and the original fingerprint.

The results achieved by the kernel, cosine, and Euclidean methods are shown in Fig.~\ref{fig:selectedResults}(a), (b), and (c), respectively. 
The results in Fig.~\ref{fig:selectedResults} indicate that all the methods produced better localization performance with a lower error when the kNN is applied with our selection strategy.
Furthermore, we can see that the neighbors have less effect on localization performance when our selection strategy is applied.
This is because the selection strategy has refined the fingerprint vector based on a list of selected beacons, in which this list of selected beacons can provide a more robust RSS value that is insensitive to the holding dynamic of a smartphone.
Hence, the similarity computation is benefited from this refined fingerprint vector and is capable of estimating the location with the best matching fingerprint rather than $k$ matching fingerprints.
Overall, the kernel method outperforms the cosine and the Euclidean methods, with localization errors less than 1~m.
In particular, kernel produced 0.64~m localization error when $k=1$, whereas cosine produced  2.50~m error, and Euclidean produced 2.52~m error.
Since the fingerprints are linearly non-separable after applying the selection strategy, cosine and Euclidean methods could not produce better location estimation as compared to the kernel method.

\begin{figure}
	\centering
	\includegraphics[width=1\columnwidth]{ 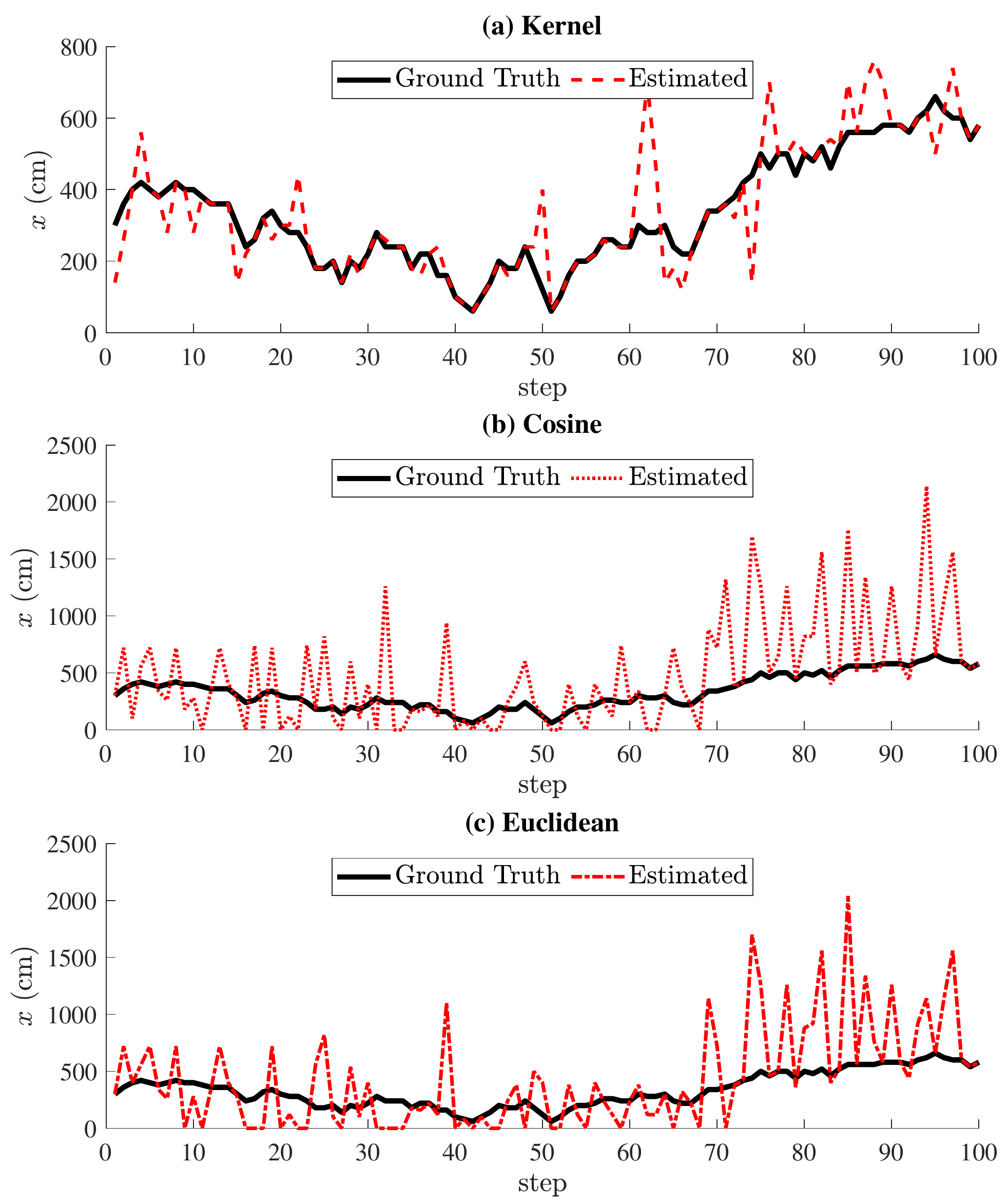}
	\vspace*{-0.5cm}
	\caption{The random path and the estimation achieved by (a) Kernel, (b) Cosine and (c) Euclidean methods.}
	\label{fig:randomXpath}
	\vspace*{-0.5cm}
\end{figure}
\subsection{Discussion}
As shown in the previous subsection, our proposed beacon selection strategy and kernel method substantially improve the localization performance in comparison to the cosine and Euclidean methods.
However, previous results only compare the localization errors produced by different methods, without explicitly indicating the estimated location given a walking path.
To get a better picture of the location estimation, we evaluated two paths by sampling the data from the testing set.
For the first path, we sampled the data along the same path we used during the on-site fingerprint survey.
For the second path, we considered a long corridor in which the $y$-coordinate is fixed, and performed a random walk sampling given a starting point.
More specifically, the user can stay in the same position, or move to the left or right from the starting point. We constrain the maximum walking length the user can move to 0.6~m by assuming the maximum possible length a normal human can walk in a single step.
We applied the best configurations for all three methods, based on the previous experimental results, to estimate the location for those two parts. The estimated walking paths produced by (a) Kernel, (b) Cosine, and (c) Euclidean methods, following the fixed path are depicted in Fig.~\ref{fig:walkpath}, whereas Fig.~\ref{fig:randomXpath} illustrates the random path.

\begin{figure}
	\centering
	\includegraphics[width=0.8\columnwidth]{ 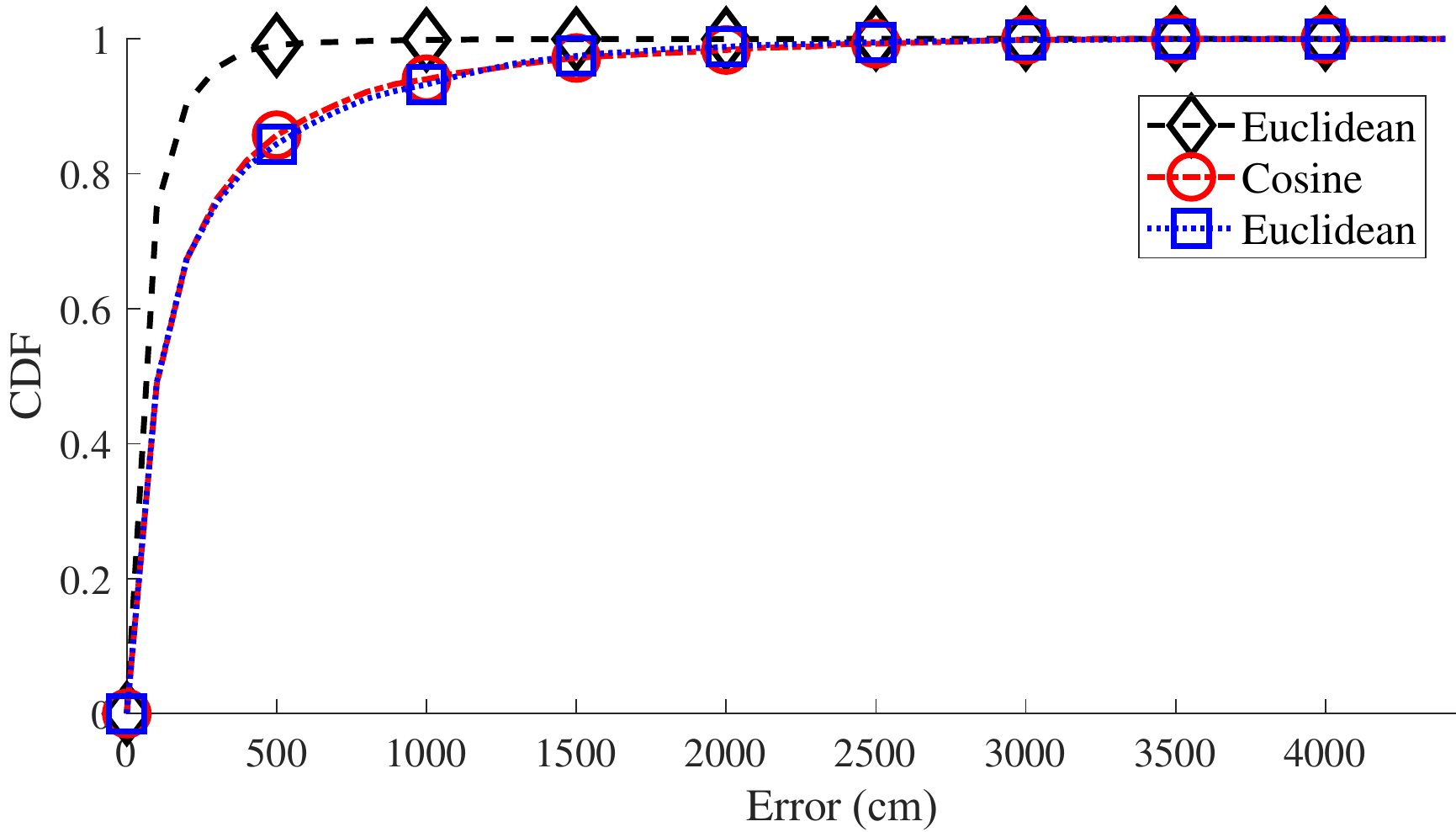}
	\vspace*{-0.3cm}
	\caption{The CDF of localization errors produced by each method.}
	\label{fig:errorCDF}
\end{figure}

Fig.~\ref{fig:walkpath} shows that the kernel method can produce almost the exact same path with small errors at the corner, whereas the Euclidean method produced a lot of errors even at the far end of the vertical path.
The superior result achieved by the kernel method confirms the feasibility of our proposed approach for any IoT applications relying on the smartphone for indoor localization.
On the other hand, Fig.~\ref{fig:randomXpath} shows that the kernel method can produce a quite robust estimation even though the user was wandering along a corridor. 
We can see that the error produced by the kernel method is relatively small compared to the cosine and the Euclidean methods.
Future work can be conducted to improve the performance of location estimation by considering the continuous trajectory information while applying our proposed method to estimate the location at every step.
Furthermore, we can also leverage the rich sensing features embedded in the smartphone to obtain hidden patterns that are useful to improve the localization performance.
Overall, our proposed beacon selection strategy with the kernel method outperformed the rest with very low localization error, as shown with the CDF plot in Fig.~\ref{fig:errorCDF}.

\section{Conclusions}
\label{sec:conclusions}
This paper highlights the importance of beacon selection before estimating the location given the observed fingerprints.
A kernel function is defined to compute the similarity between the observation vector and fingerprints in the high-dimensional space without literally visiting the space.
While our experimental results verified the superior performance of our proposed approach over existing approaches, further work can be conducted to enhance the overall performance from various aspects.
First, a hybrid selection approach to the beacon selection strategy can be developed so that the beacon selection strategy is not only performed during offline fingerprinting but adaptively during online estimation.
This is critical to deal with the case when some of the beacons have stopped functioning due to the dead battery.
Second, a classification method can be applied to learn the decision boundary that separates the fingerprints rather than having to visit all the fingerprints for similarity computation.
Lastly, we can also exploit the inertial sensors embedded in the smartphone to estimate the smartphone's motion before estimating the location using the fingerprint.






%
%
%

\bibliographystyle{IEEEtran}
\bibliography{references}

\begin{IEEEbiography}[{\includegraphics[width=1in,height=1.25in,clip,keepaspectratio]{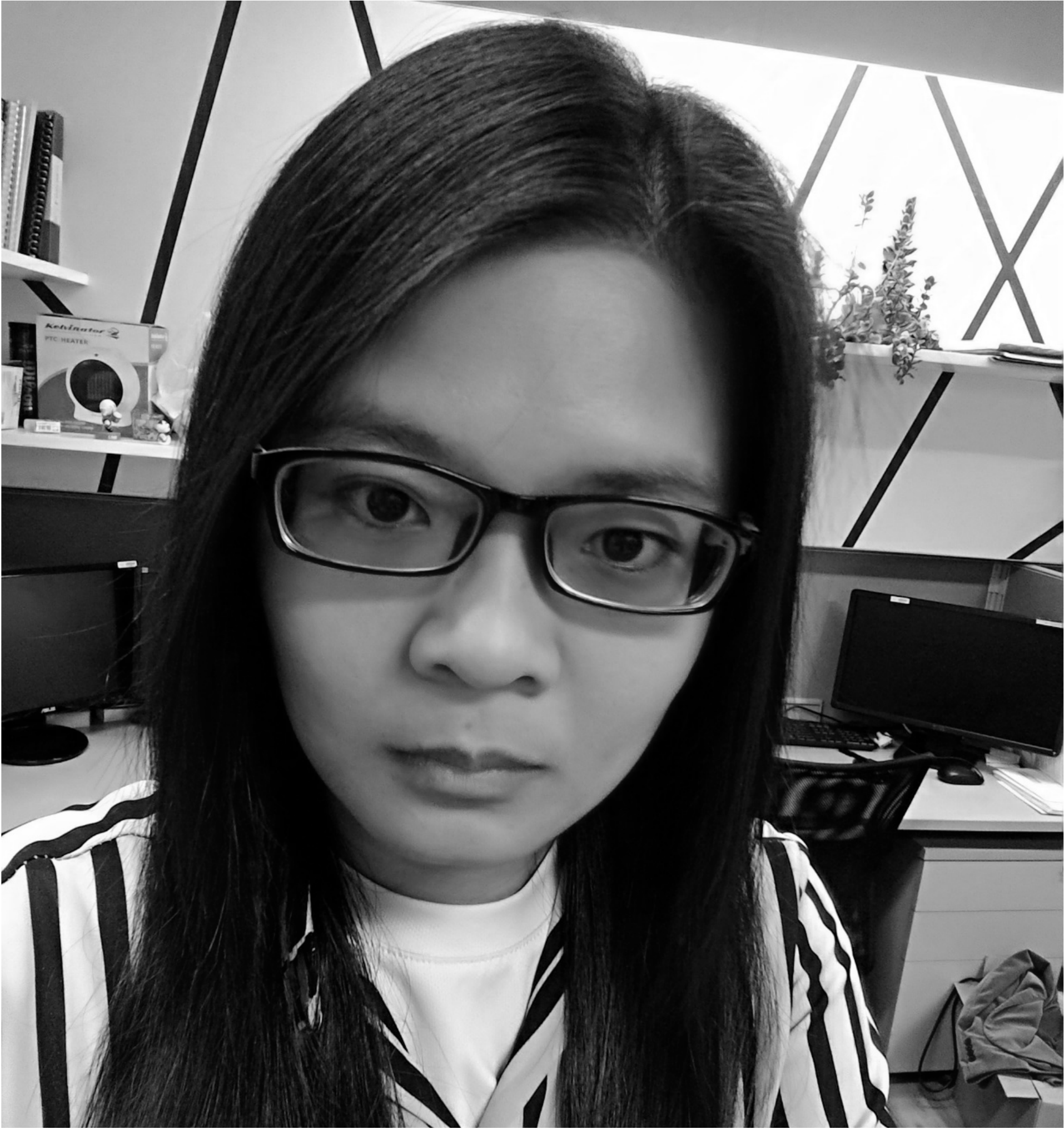}}]{Pai Chet Ng}
	is currently a Post-doctoral Researcher with the University of Toronto. She obtained her Ph.D. degress in electronic and computer engineering from the Hong Kong University of Science and Technology. Her research interests include Proximity-based Sensing and Networking,  Physiological Signals with Mobile and Wearable Devices, and IoT Systems with Artificial Intelligence.
\end{IEEEbiography}

\vspace*{-1cm}
\begin{IEEEbiography}[{\includegraphics[width=1in,height=1.25in,clip,keepaspectratio]{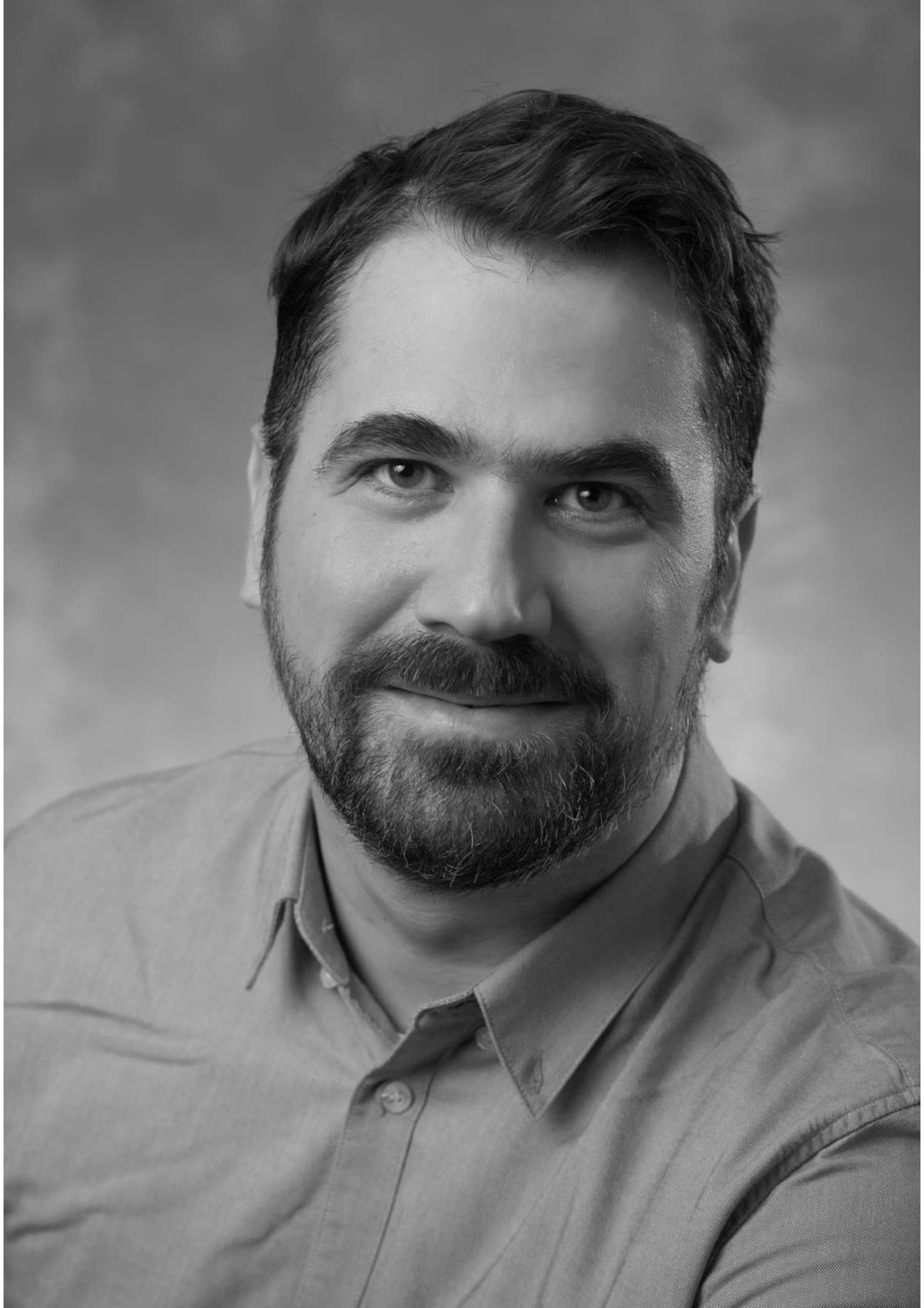}}]{Petros Spachos} received the Diploma in electronic and computer engineering degree from the Technical University of Crete, Chania, Greece, and the M.A.Sc. and Ph.D. degrees in electrical and computer engineering from the University of Toronto, ON, Canada. He is currently an Associate Professor with the School of Engineering, University of Guelph, ON, Canada.  He is also a registered professional engineer in Ontario. His research interests include experimental wireless networking and mobile computing with a focus on wireless sensor networks, smart cities, and the Internet of Things.

\end{IEEEbiography}
 
\vspace*{-1cm}
\begin{IEEEbiography}[{\includegraphics[width=1in,height=1.25in,clip,keepaspectratio]{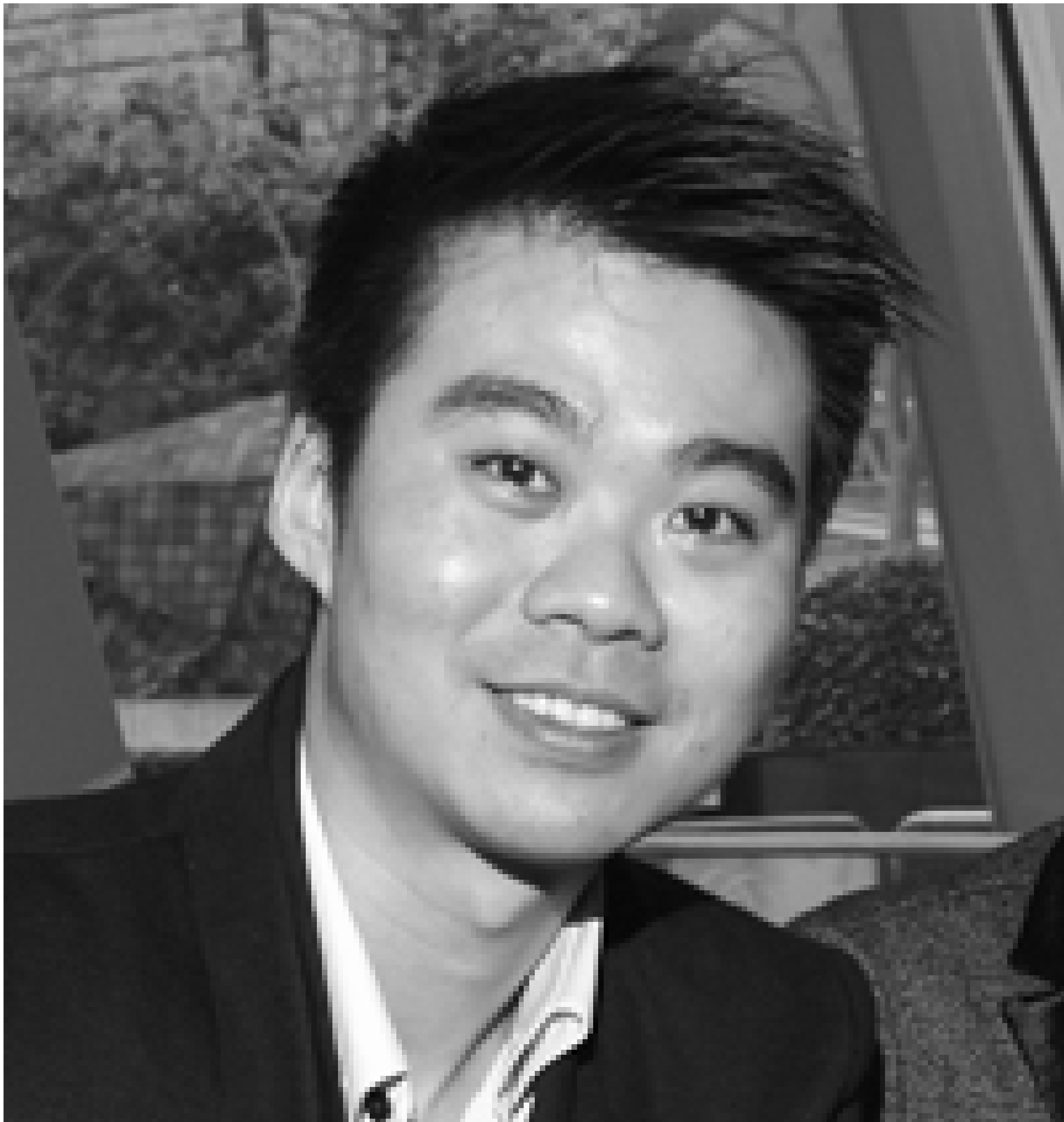}}]{James She}
is an Associate Professor in the Division of Information and Computing Technology, College of Science and Engineering at Hamad Bin Khalifa University, Qatar. His current research interests include IoT and Multimedia, especially with the uses of emerging machine learning and AI technologies. He is also an adjunct faculty member at Hong Kong University of Science and Technology, Hong Kong. 
\end{IEEEbiography}

\vspace*{-1cm}
\begin{IEEEbiography}[{\includegraphics[width=1in,height=1.25in,clip,keepaspectratio]{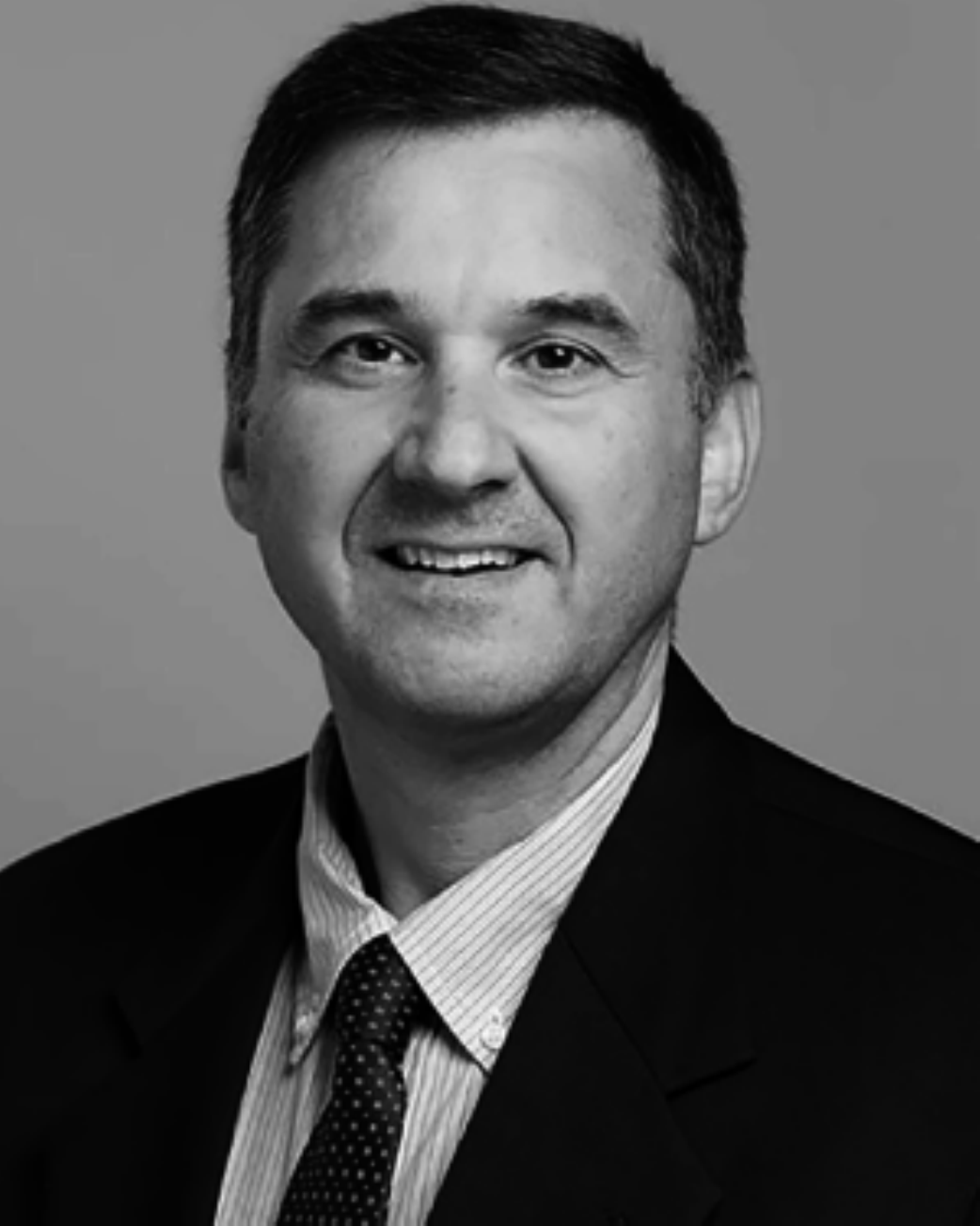}}]{Konstantinos N. Plataniotis} is a Professor and the Bell Canada Chair in Multimedia at the University of Toronto. His research interests are in the areas of machine learning and signal processing, and their applications in imaging systems, communications, and knowledge media design systems. Dr. Plataniotis is a Fellow of the Engineering Institute of Canada, Fellow of the Canadian Academy of Engineering / L’ Academie Canadienne Du Genie, and a registered professional engineer in Ontario. He is the General Co-Chair of the 2027 IEEE International Conference on Acoustics, Speech and Signal Processing (ICASSP2027).
\end{IEEEbiography}

\end{document}